\documentclass[12pt]{iopart}
\usepackage{iopams}
\usepackage{amssymb}
\usepackage{cuted}
\usepackage{textcomp}
\usepackage{graphicx}
\usepackage{grffile}
\usepackage{color}
\usepackage[caption=false]{subfig}
\usepackage{bm}
\usepackage{tabularx}
\definecolor{Green}{rgb}{0.000000,0.392157,0.000000}
\usepackage{placeins}
\newcommand{\stl}[1]{ \mbox{ $ \stackrel{ \rule[-1.5pt]{0.5pt}{0.65ex} \overline{\hphantom{\mbox{$\displaystyle #1$}}} \rule[-1.5pt]{0.5pt}{0.65ex} }{#1} $ } }
\renewcommand{\vec}[1]{\bm{#1}}
\usepackage{mathrsfs}
\usepackage{soul}

\begin{document}


\title[Binary mixtures of rod--like colloids under shear]{Binary mixtures of rod--like colloids under shear: microscopically--based equilibrium theory and order--parameter dynamics}
\author{Rodrigo~Lugo-Fr\'ias and Sabine~H.~L.~Klapp}
\address{Institute f\"ur Theoretische Physik, Technische Universit\"at Berlin, Hardenbergstrasse 36, 10623 Berlin, Germany}
\ead{klapp@physik.tu-berlin.de}


\begin{abstract}
This paper is concerned with the dynamics of a binary mixture of rod--like, repulsive colloidal particles driven out of equilibrium by means of a steady shear flow (Couette geometry).
To this end we first derive, starting from a microscopic density functional in Parsons--Lee approximation, 
a mesoscopic free energy functional whose main variables are the orientational order parameter tensors. 
Based on this mesoscopic functional we then explore the stability of isotropic and nematic equilibrium phases in terms of composition and rod lengths.
Second, by combining the equilibrium theory with the Doi--Hess approach for the order parameter dynamics under shear, we investigate 
the orientational dynamics of binary mixtures for a range of shear rates and coupling parameters.
We find a variety of dynamical states, including synchronized oscillatory states of the two components, but also
symmetry breaking behavior where the components display different in--plane oscillatory states.
\end{abstract}

%
\vspace{2pc}


\section{Introduction}
Mixtures of rod--like colloidal particles are found in many natural systems, e.g., in cells and in the blood,
as well as in various industrial applications such as paints and in food science. Typically, these colloidal systems are dominated by the repulsive, anisotropic interactions stemming from the particle's
non--spherical shape. 
Already in thermodynamic equilibrium, such mixtures
display a very rich, entropy--driven phase behavior~\cite{Kooij2000, Purdy2005, Varga2005, Brinke2008, Westen2014}, characterized by the
interplay of orientational ordering, (depletion--induced) condensation and demixing. A recent example is an experimental study of polydisperse clay rods which show
an isotropic--nematic phase transition on increasing concentration~\cite{Woolston2015}. From the theoretical side, 
such phase diagrams have been studied, e.g., by Onsager theory~\cite{Onsager1949}
and by more general free energy functionals~\cite{Parsons1979, Lee1987, Tarazona1989} in the framework of classical density functional theory (DFT), as well as
by computer simulations~\cite{Bolhuis1997, Cuetos2007}.

While the equilibrium behavior is already challenging, a further level of complexity is reached when considering such mixtures under shear flow~\cite{Loewen2008, Loewen2012, Guu2013}. 
The current interest in this topic has basically two reasons: first, investigating the
impact of (shear--) flow is crucial for understanding the system's rheological behavior, an aspect which is particularly important from the application side.
Second, shear flow is a fundamental example of driving a soft--matter system out of equilibrium. Indeed, already for one--component systems of colloidal rods,
shear can have profound consequences. Examples are the shear--induced shift of the isotropic--nematic transition and the associated
stabilization of the nematic phase~\cite{Olmsted1991, Dogic2007}, and the occurrence of spatial symmetry breaking 
related to shear--banding~\cite{Fielding2003, Fielding2006, Olmsted2008}. Moreover, even under steady shear the rod--like particles can display spontaneous time--dependent oscillations, a prime
example being the tumbling motion (characterized by full rotations of the particle axes). Indeed, tumbling has been observed in 
experiments~\cite{Lettinga2005} as well as in particle--resolved computer simulations~\cite{Ripoll2008, Tao2009}.

Another well--established route to explore the orientational dynamics under shear are 
mesoscopic (continuum) equations, such as the Doi--Hess approach
\cite{Ericksen1961, Leslie1968, Hess1975, Hess1976, Doi1978, Pardowitz1980}, describing the evolution of an appropriate order parameter in time.
For one--component systems of rods with head--tail symmetry, this mesoscopic order parameter is the so--called $\mathbf{Q}$--tensor, a second--rank tensor which corresponds to the lowest non--vanishing moment of the 
one--body orientational distribution function $f=f(\bm{\bm{\omega}})$ (where $\bm{\bm{\omega}} = (\theta,\phi)$ contains 
the set of Euler angles describing the particle's orientation~\cite{Gennes1993, Hess2015}).
Within the Doi--Hess approach~\cite{Hess1975, Hess1976, Doi1978}, the time dependence of $\mathbf{Q}$ is described by 
a competition between flow--induced perturbations, on the one hand, and a relaxational contribution stemming from the (functional) derivative of the phenomenological Landau--de Gennes
(LdG)  free energy~\cite{Gennes1993}, on the other hand. The underlying equilibrium behavior is then determined by the coefficients of the powers of the order parameter in the LdG free energy.
Usually, these coefficients are constants which do not depend on a specific microscopic Hamiltonian, and are thus, independent of e.g., the aspect ratio of the particles; rather
one uses the LdG functional as an universal free energy expression yielding an isotropic--nematic (I--N) transition.
The resulting, nonlinear equations of motion indeed generate complex dynamical behavior characterized by different (regular) oscillatory states~\cite{Rienaecker1999, Rienaecker2000, Strehober2013}, chaotic behavior~\cite{Rienaecker2002} and spatio--temporal patterns~\cite{Heidenreich2006, Das2005}.

In the present paper we employ mesoscopic equations inspired by the Doi--Hess approach to investigate the shear--induced dynamics of binary rod mixtures. However,
we modify the original approach by replacing the standard LdG free energy functional (for mixtures) by a free energy functional
derived from classical DFT. In this way, our theory incorporates microscopic information, particularly the aspect ratios and the molar fractions characterizing the two species in the mixtures. 
 
Classical DFT is a powerful tool to study inhomogeneous fluids near interfaces and in other external potentials~\cite{Evans1979, Evans1992}
and has nowadays a wide range of applications both for sphere--like systems and for systems of shape--anisotropic particles. 
The key idea of DFT is to minimize a (free energy) functional of the singlet density $\rho(\mathbf{r},\bm{\bm{\omega}})$~\cite{Tarazona1989, Tarazona2008}. 
Several studies have been devoted to construct free energy functionals for homogeneous states of anisotropic fluids,
such that $\rho(\mathbf{r},\bm{\bm{\omega}}) = \rho f(\bm{\bm{\omega}})$ (where $\rho = N/V$ is the number density). Here one main goal
was to describe the I--N phase transition~\cite{Singh2000, Vink2005}. Of special importance for the present study is
the work of Malijevsky {\em et al.}~\cite{Malijevsky2008}, who developed a DFT for the description of spatial homogeneous phases (isotropic and nematic) of hard--body fluid mixtures by
generalizing the ideas of Parsons~\cite{Parsons1979} and Lee~\cite{Lee1987} for pure component systems.

Here we use this multicomponent Parsons--Lee theory~\cite{Malijevsky2008} as a starting point to construct an (approximate) free energy functional containing 
powers of the tensorial order parameters $\mathbf{Q}^i$, where $i=A,B$ labels the two components of the mixture. The coefficients in this expansion contain the microscopic information
mentioned before. We then combine our equilibrium theory with a straightforward extension of the dynamical Doi--Hess equations for a binary mixture to study the full order parameter dynamics of the system under (Couette) shear flow.
By combining microscopic (DFT) and mesoscopic "ingredients", we consider our approach as a contribution to a scale--bridging description of complex colloidal mixtures far from equilibrium.
To some extent, our work is related to recent derivations of phase--field--crystal models (which include gradient terms)
starting from microscopic functionals~\cite{Wittkowski2010, Menzel2015}.

The paper is organized as follows. In section 2 we first define in detail the hard--body binary mixture and present the microscopic (Parsons--Lee) free energy functional. 
We then perform an expansion into $\mathbf{Q}$--tensors, yielding a "mesoscopic" functional representing the orientational part of the full free energy up to fourth order 
in the order parameters.
In sections 2.3 and 2.4, we then specialize to mixtures of hard spherocylinders in the semi--dilute regime and investigate the stability of isotropic and nematic states (for fixed densities).
Section 3.1 is devoted to the explicit equations of motion in the framework of the Doi--Hess theory. In section 3.3. we present numerical results
including full non--equilibrium state diagrams portraying the dynamic behavior of the orientational order parameters. The paper finishes with concluding remarks and an outlook in section 4. 
Some technical details are given in an appendix.


\section{Equilibrium Theory}
\label{Sec:equilibrium}

We consider a binary mixture of hard, uniaxial particles with repulsive pair interactions dependent on their orientations. The two species are labeled by indices $i$, $j$, which take
values $A$ or $B$.
The pair potential between a particle "$1$" of component $i$ and a particle "$2$" of component $j$ is then given by (see, e.g.,~\cite{Malijevsky2008})
\begin{equation}
\label{Eq:01}
  u_{ij}(\vec{r}_{12},{\bm{\omega}}_1,{\bm{\omega}}_2)=\cases{\infty & $r_{12} < d_{ij}({\bm{\omega}}_1,{\bm{\omega}}_2,{\bm{\omega}}_{12})$\\
0 & otherwise \\}
\end{equation}
where $d_{ij}({\bm{\omega}}_1,{\bm{\omega}}_2,{\bm{\omega}}_{12})$ is the orientational dependent distance of closest approach of the two particles,
$r_{12}$ is the magnitude of the center--to--center vector $\mathbf{r}_{12} = r_{12}\bm{\omega}_{12}$,
$\bm{\omega}_1=(\theta_1, \varphi_1)$ and $\bm{\omega}_2=(\theta_2, \varphi_2)$ are the Euler angles of the two particles,
and ${\bm{\omega}}_{12}= (\theta_{12}, \varphi_{12})$ describes the orientation of $\mathbf{r}_{12}$.

\subsection{Microscopic free energy functional}

Within the formalism of classical DFT, the total Helmholtz free energy 
is written as a sum of ideal and excess contributions, such that $\mathcal{F} = \mathcal{F}_{id} + \mathcal{F}_{ex}$~\cite{Evans1979, Evans1992}. This quantity is a functional of the one--particle densities 
$\rho_i(\vec{r},{\bm{\omega}})$. Here we focus on spatial homogeneous phases such that $\rho_i(\vec{r},{\bm{\omega}}) = \rho_i f_i(\bm{\bm{\omega}})$, where
$\rho_i = N_i/V$ is the number density of species $i$ (with $V$ being the volume) and
$f_i({\bm{\omega}})$ is the orientational distribution function (ODF), which is normalized such that
$\int d\bm{\omega} f_i(\bm{\omega})=1$. This ansatz allows us to describe isotropic and nematic phases of the system.

With this specification, the ideal part of the free energy functional can be written as 
\begin{equation}
\label{Eq:02}
 \frac{\beta \mathcal{F}_{id}}{V} [\{\rho_i\}]= \sum_{i=A,B} \rho_i \left( \ln(\rho_i \Lambda_i^3) - 1 + \sigma[f_i] \right)\,,
\end{equation}
where $\beta = 1/k_B T$ ($T$ being the temperature and $k_B$ Boltzmann's constant),
$\Lambda_i$ is the thermal de Broglie wavelength of each component, and $\sigma[f_i]$ is the orientational entropy given by
\begin{equation}
\label{Eq:03}
 \sigma[f_i] = \int d{\bm{\omega}} f_i({\bm{\omega}}) \ln(4\pi f_i({\bm{\omega}})),
\end{equation}
where $\int d\bm{\omega}=\int_{-1}^{1}d\cos\theta\int_{0}^{2\pi}d\varphi$.
The quantity $\sigma_i$ is zero in the isotropic state, where $f_i(\bm{\omega})=f_0=(4\pi)^{-1}$ and becomes positive in the nematic state. It thus reflects the loss of orientational entropy
due to ordering.

Regarding the excess part of the free energy, we here employ a generalization of the so--called Parsons--Lee theory~\cite{Parsons1979, Lee1987} to mixtures, as suggested by
Malijevsky {\em et al.}~\cite{Malijevsky2008}. Briefly, this approach is derived by starting from the virial expression for the pressure, which involves (apart from the singlet densities)
the spatial derivative of the pair potential, $u_{ij}$ and the pair correlation function $g_{ij}$. 
Noting that $u_{ij}$ can be re--expressed in terms of a hard--sphere (HS) potential with scaled distance $r_{12}/d_{ij}(\bm{\omega}_1,\bm{\omega}_2,\bm{\omega}_{12}$), the same assumption
is made for the correlation function, i.e., $g_{ij}=g_{ij}^{HS}(r_{12}/d_{ij})$ (note that this is an approximation). By integrating over the total density $\rho=\sum_{i}\rho_i$ (and noting
that $\partial u_{ij}^{HS}(y)/\partial y$ is proportional to a delta function in space), one obtains
the excess free energy
\begin{eqnarray}
\label{Eq:04}
\fl \frac{\beta \mathcal{F}_{ex}}{V}[\{\rho_i\}] = \frac{1}{2}&\sum_{i =A,B} \sum_{j=A,B} \rho_i \rho_j G_{ij} \int d{\bm{\omega}}_1  \int d{\bm{\omega}}_2  V_{ij}^{exc} ({\bm{\omega}}_1,{\bm{\omega}}_2)f_i({\bm{\omega}}_1)f_j({\bm{\omega}}_2)\,.
\end{eqnarray}
In~\eref{Eq:04}, $V_{ij}^{exc}$ is the excluded volume between particles $1$ and $2$, which is related to the contact distance via
\begin{eqnarray}
\label{Eq:05}
V_{ij}^{exc}(\bm{\omega}_1,\bm{\omega}_2)=\frac{1}{3}\int d \bm{\omega}_{12} d_{ij}^3(\bm{\omega}_1,\bm{\omega}_2,\bm{\omega}_{12}).
\end{eqnarray}
Further, the quantities $G_{ij}$ are averages (in terms of the total density) of the contact values of the HS pair correlation functions, $g_{ij}^{HS}(1_{+})$, in the considered density range.

As discussed in ~\cite{Malijevsky2008}, the free energy given in~\eref{Eq:04} reduces to 
the more familiar second--virial theory of Onsager~\cite{Onsager1949} in the limit of low densities. Then, one has $g_{ij}^{HS}(1_{+})\rightarrow 1$, and therefore $G_{ij}=1$ for all combinations
$i,j$. Although the Onsager theory provides an intuitive approach regarding the effects related to the geometry of the particles, it is known to become inaccurate for particles of finite length at the densities close to orientational phase transitions~\cite{Bolhuis1997, Cuetos2007}. In this respect the generalized Parsons--Lee approach
provides a significant improvement, as demonstrated by comparison with Monte Carlo simulations of hard Gaussian molecules with different elongations and concentrations~\cite{Zhou2004}.

\subsection{Mesoscopic free energy functional}

The aim of this section is to rewrite the microscopic free energy functional consisting of the ideal and excess parts given in equations~\eref{Eq:02} and~\eref{Eq:04}, respectively,
in terms of ''mesoscopic" orientational order parameters, specifically, the tensors, $\mathbf{Q}^i_{(k)},\,i=A,B$, of rank $k$. These tensors are defined as~\cite{Hess2015}
\begin{equation}
\label{Eq:06}
\mathbf{Q}^i_{(k)} = \sqrt{\frac{(2k+1)!!}{k!}} \Big\langle\stl{{\bm{\omega}}^{(k)}}\Big\rangle_i\,,
\end{equation}
where $\bm{\omega}$ is a unit vector with the orientation $\bm{\omega}=(\theta,\varphi)$,
$\langle{\cdots}\rangle_i=\int d\bm{\omega}_i f(\bm{\omega}_i)\cdots$ is an average involving the orientational distribution function,
and the notation $\stl{\mathbf{A}}$ denotes the symmetric traceless part of the tensor ${\mathbf{A}}$. Here, $\mathbf{A}={\bm{\omega}}^{(k)}$ is 
a $k$--fold tensorial product of the form
${{\bm{\omega}}\otimes \cdots \otimes {\bm{\omega}}}$ (e.g., for the case $k=2$ this yields
$({\bm{\omega}}^{(2)})_{mn}=({\bm{\omega}})_m({\bm{\omega}})_n$).

For the present systems, which involve uniaxial particles with head--tail symmetry, it is sufficient to consider tensors with even $k$. Within this subset, 
the tensors of rank $k=2$ and $k=4$ are of particular importance since they are related
to physical phenomena such as birefringence and fluorescence~\cite{Gennes1993, Pardowitz1980}. Tensors with odd values of $k$ occur in the description
of systems with additional degrees of freedom such as rod--like ferrofluids~\cite{Kroeger2008, Grandner2007}.

\subsubsection{Ideal contribution}

The first task is to rewrite the orientational entropy, given in~\eref{Eq:03}. Here we follow earlier approaches ~\cite{Pardowitz1980, Grandner2007}, where
the one--particle ODF was expressed as $f_i({\bm{\omega}}) = f_0\left(1 + \alpha_i({\bm{\omega}})\right)$, with $\alpha({\bm{\omega}})$ being a small deviation from the isotropic state, 
$f_0 = 1/4\pi$. Inserting this ansatz into~\eref{Eq:03} and performing a Taylor expansion of the logarithm yields~\cite{Pardowitz1980,McCourt1991}
\begin{equation}
\label{Eq:07}
\sigma[f_i] = \sum_{n=2}^\infty \frac{(-1)^n}{n(n-1)}\langle  \alpha_i^n({\bm{\omega}}) \rangle_0 \,,
\end{equation}
where $\langle{\cdots}\rangle_0$ denotes the orientational average evaluated with $f_0$. Due to normalization of the ODF the term $n=1$ is zero and thus it is disregarded. Next, one assumes
that the function $\alpha({\bm{\omega}})$ can be expanded in terms of the tensorial quantities $\left((2k+1)!!/k!\right)\stl{{\bm{\omega}}^{(k)}}$ [see~\eref{Eq:06}], where
the prefactors are determined by their orientational averages, $\mathbf{Q}^i_{(k)}$. Following ~\cite{Pardowitz1980,McCourt1991},
we restrict this expansion to terms involving $k=2$ and $k=4$. Inserting the resulting {\em ansatz} for $\alpha_i$ into~\eref{Eq:07} and retaining terms up to fourth power
in $\mathbf{Q}^i_{(2)}$ and up to second power in $\mathbf{Q}^i_{(4)}$, one obtains
\begin{eqnarray}
\label{Eq:08}
\fl \sigma\left(\mathbf{Q}_{(2)}^i,\mathbf{Q}_{(4)}^i\right) =  \frac{1}{2}&\mathbf{Q}_{(2)}^i \odot \mathbf{Q}_{(2)}^i- \frac{\sqrt{30}}{21}\mathbf{Q}_{(2)}^i \cdot \mathbf{Q}_{(2)}^i \odot \mathbf{Q}_{(2)}^i\nonumber \\
& + \frac{5}{28}\left[\mathbf{Q}_{(2)}^i \odot \mathbf{Q}_{(2)}^i\right]^2 + \frac{1}{2}\mathbf{Q}_{(4)}^i \odot \mathbf{Q}_{(4)}^i.
\end{eqnarray}
In~\eref{Eq:08}, the product $\mathbf{A}_{(k)} \cdot \mathbf{B}_{(k)}$ yields a tensor of rank $k$ (with components
$A_{ij \ldots mn}B_{nm \ldots kl} = C_{ij \ldots kl}$ (where we used the Einstein convention). 
Further, $\mathbf{A}_{(k)} \odot \mathbf{B}_{(k)} = A_{ij \ldots kl}B_{ij \cdots kl} =c$ indicates the maximal possible contraction,
that is, $c$ is a scalar.

\subsubsection{Excess contribution}

We now turn to the reformulation of the excess free energy as functions of the $\mathbf{Q}^i_{(k)}$. To our knowledge, this problem has not been addressed in the literature so far.
As a starting point we expand the excluded volume of the hard--core particles appearing in~\eref{Eq:04} in terms of the Legendre polynomials
(following~\cite{Malijevsky2008, Vroege1992} and references therein),
\begin{equation}
\label{Eq:09}
V^{exc}_{ij} ({\bm{\omega}}_1,{\bm{\omega}}_2) = \sum_{k=0}^\infty V_k^{ij} P_k(\cos\gamma),
\end{equation}
where $\gamma$ is the angle between ${\bm{\omega}}_1$ and ${\bm{\omega}}_2$, and the coefficients $V_k^{ij}$ are calculated from the relation
\begin{equation}
\label{Eq:10}
V_k^{ij} = \frac{2k+1}{2} \int_{-1}^{1}d\cos\gamma P_k(\cos\gamma) V^{exc}_{ij}(\gamma).
\end{equation}
Next, we replace $P_k$ in~\eref{Eq:09} by using the addition theorem for spherical harmonics~\cite{Gray1984}, yielding
\begin{equation}
\label{Eq:11}
V^{exc}_{ij} ({\bm{\omega}}_1,{\bm{\omega}}_2) = \sum_{k=0}^\infty \frac{4\pi V_k^{ij} }{2k + 1} \sum_{m=-k}^{k} Y^*_{km}({\bm{\omega}}_1)Y_{km}({\bm{\omega}}_2)\,.
\end{equation}
The sum on the right side of~\eref{Eq:11} can be rewritten as a 
full scalar product of the two tensors $\stl{\bm{\omega}^{(k)}}$~\cite{Gray1984}. 
One obtains
\begin{equation}
\label{Eq:12}
V^{exc}_{ij} ({\bm{\omega}}_1,{\bm{\omega}}_2) = \sum_{k=0}^\infty V_k^{ij} \frac{(2k+1)!!}{(2k + 1)k!} \stl{{\bm{\omega}}_1^{(k)}} \odot \stl{{\bm{\omega}}_2^{(k)}}\,.
\end{equation}
Inserting~\eref{Eq:12} into~\eref{Eq:04}, noting that the angular integrals (weighted by the ODFs) in~\eref{Eq:04} lead to orientational averages,
and using the definition of the tensorial order parameter [see~\eref{Eq:06}] one finds
\begin{equation}
\label{Eq:13}
 \frac{\beta \mathcal{F}_{exc}}{V} = \frac{1}{2}\sum_{i,j} \rho_i \rho_j G_{ij} \sum_{k=0}^\infty \frac{V_k^{ij}}{2k + 1} \mathbf{Q}^i_{(k)} \odot \mathbf{Q}^j_{(k)}\,.
\end{equation}
Equation~\eref{Eq:13} provides the excess free energy as a function of order parameter tensors. It still depends
on the microscopic parameters of the mixture,
that is, the geometry of the particles (defining the excluded volume) and the densities. 

\subsubsection{Full free energy}

We now turn to the full free energy containing ideal and excess parts, $\mathcal{F}=\mathcal{F}_{id}+\mathcal{F}_{ex}$. Restricting to terms involving $2nd$-- and $4th$-- rank tensors we have
 \begin{equation}
\label{Eq:14}
 \frac{\beta F_{id}[\{\mathbf{Q}\}]} {V}=
 \sum_{i=A,B} \left[\mathcal{F}_{00}^i + \rho_i \sigma\left(\mathbf{Q}_{(2)}^i,\mathbf{Q}_{(4)}^i\right)\right]\,,
\end{equation}
where $\mathcal{F}_{00}^i =\rho_i(\ln(\rho_i \Lambda_i) - 1)$, $\sigma$ is given in~\eref{Eq:08}, and the notation
$[\{\mathbf{Q}\}]$ on the left side indicates the dependence of the (ideal) free energy on all four order parameter tensors ($k=2,4$, $i=A,B$). Further, we obtain from~\eref{Eq:13}
\begin{eqnarray}
\label{Eq:15}
\fl \frac{\beta \mathcal{F}_{ex}[\{\mathbf{Q}\}]}{V} = \frac{1}{2}\sum_{i,j} \rho_i \rho_j G_{ij} \left[ V_0^{ij} +\frac{V_2^{ij}}{5}(\mathbf{Q}_{(2)}^i \odot \mathbf{Q}_{(2)}^j) + \frac{V_4^{ij}}{9}  (\mathbf{Q}_{(4)}^i \odot \mathbf{Q}_{(4)}^j) \right]\,.
\end{eqnarray}
These expressions can be further simplified by using a ''closure" relation expressing the $4th$--rank tensor with the one of second rank. For uniaxial systems characterized
by a (nematic) director $\mathbf{n}$~\cite{Kroeger2008, Kroeger2003},
there exists the {\em exact} relation $S_2^2 \mathbf{Q}_{(4)} = S_4 \stl{\mathbf{Q}_{(2)} \mathbf{Q}_{(2)}}$, where $S_2 = \langle P_2(\bm{\omega} \cdot \mathbf{n}) \rangle$ and $S_4 = \langle P_4(\bm{\omega} \cdot \mathbf{n}) \rangle$. In strongly aligned systems one has $S_4\approx S_2^2\approx 1$. In that limit, the closure thus takes the form
$\mathbf{Q}_{(4)} \equiv \stl{\mathbf{Q}_{(2)} \mathbf{Q}_{(2)}}$. 

Since we can now express all quantities in terms of the second--rank tensor, we ease notation and set $\mathbf{Q}^i=\mathbf{Q}^i_{(2)}$ ($i=A,B$). 
Further, we rewrite the notation for the tensor products appearing in~\eref{Eq:15} into the form commonly used for second--rank tensors, that is
$\mathbf{Q}^i \odot \mathbf{Q}^i=\mathbf{Q}^i:\mathbf{Q}^i= \Tr(\mathbf{Q}\cdot\mathbf{Q}) = \sum_{mn} Q_{mn}Q_{mn}$. 
Adding ideal and excess contributions of the free energy, we obtain
\begin{eqnarray}
\label{Eq:16}
\fl \frac{\beta \mathcal{F}(\{\mathbf{Q}\})}{V}=
 \sum_{i=A,B} &\left\lbrace  \mathcal{F}_0^i + A_i (\mathbf{Q}^i:\mathbf{Q}^i) - B_i \Tr(\mathbf{Q}^i \cdot \mathbf{Q}^i \cdot \mathbf{Q}^i) + C_i (\mathbf{Q}^i:\mathbf{Q}^i)^2) \right\rbrace \nonumber \\
&+ \mathcal{F}_0^{AB} + A_{AB} (\mathbf{Q}^A : \mathbf{Q}^B) + C_{AB}(\mathbf{Q}^A : \mathbf{Q}^B)^2 \,,
\end{eqnarray}
where the coefficients are functions of the number densities, the factors $G_{ij}$ and the coefficients of the excluded volume. Explicitly, they are given by
\numparts
\begin{eqnarray}
\label{Eq:17a}
\mathcal{F}_0^i    &= \rho_i(\ln(\rho_i \Lambda_i) - 1) + \frac{1}{2} \rho_i^2 G_{ii} V_0^{ii}\,,\\
\label{Eq:17b}
A_i      &= \frac{1}{2}\rho_i + \frac{1}{10} \rho_i^2 G_{ii} V_2^{ii}\,,\\
\label{Eq:17c}
B_i      &= \frac{\sqrt{30}}{21}\rho_i\,,\\
\label{Eq:17d}
C_i      &= \frac{19}{28}\rho_i + \frac{1}{18}\rho_i^2 G_{ii} V_4^{ii}\,,\\
\label{Eq:17e}
\mathcal{F}_0^{AB} &= \rho_A \rho_B G_{AB} V_0^{AB}\,,\\
\label{Eq:17f}
A_{AB}   &= \frac{1}{5}  \rho_A \rho_B G_{AB} V_2^{AB}\,,\\
\label{Eq:17g}
C_{AB}   &= \frac{1}{9}  \rho_A \rho_B G_{AB} V_4^{AB}\,.
\end{eqnarray}
\endnumparts
As seen from~\eref{Eq:16}, the ''mesoscopic'' free energy derived here has the standard Landau--de Gennes (LdG) form~\cite{Gennes1993}
in the sense that it involves an expansion into order parameters. Contrary to standard LdG theory, however, 
the coefficients are related to microscopic properties of the system. We also note that 
the free energy does not depend explicitly on the temperature, which reflects the hard--core character of interactions. Thus, the ordering behavior of the resulting
''lyotropic'' system is essentially controlled by the concentration of the system. 

\subsection{Specialization to mixtures of hard spherocylinders}
The derivation presented so far can be applied to binary mixtures of {\em any} hard--core particles which are uniaxial
and have head--tail symmetry. In the following we specify on systems of hard spherocylinders.
Hard spherocylinders are characterized by two geometrical parameters, their length $l_i$ and their diameter $d_i$, defining their aspect ratio $\kappa_i = l_i/d_i$.
The resulting particle volume is $v_i = ({\pi}/{12})(3\kappa_i + 2) d_i^3$. The excluded volume $V_{exc}(\gamma)$ [defined in~\eref{Eq:05}] of two spherocylinders was
calculated by Onsager in 1949~\cite{Onsager1949}; a modern derivation has been given in the framework of general expressions for excluded volumes between
hard sphero--zonotopes (see~\cite{Mulder2005}). The result is
\begin{eqnarray}
\label{Eq:18}
\fl V_{exc}(\gamma) = \frac{\pi}{4} (\kappa_i d_i^3 + \kappa_j d_j^3) + \frac{\pi}{4}d_id_j(d_i + d_j)\sin\gamma + \frac{\pi}{4} (\kappa_i d_i d_j^2 + \kappa_j d_j d_i^2) |\cos\gamma| \nonumber \\
 + \kappa_i\kappa_jd_id_j (d_i + d_j) \sin\gamma + (\kappa_i d_i + \kappa_j d_j) d_i d_j E(\sin\gamma) \,,
\end{eqnarray}
where $E(\sin\gamma)$ denotes the complete elliptic integral of the second kind. Within the original Onsager theory, only terms up to linear order in
$\sin\gamma$ are considered; that is, the terms involving $|\cos\gamma|$ and $E(\sin\gamma)$ are neglected. In the present work
we keep the {\em full} expression~\eref{Eq:18}. 
To calculate the resulting expansion coefficients
 $V_k^{ij}$ which are defined in~\eref{Eq:10} and which determine the coefficients appearing in the free energy [see~\eref{Eq:17a}--\eref{Eq:17g}], we perform a series expansion as described
 in~\ref{ExcludedVolume}. In this way we obtain analytic expressions for the coefficients of orders $k=0,2,4$.
 
In the following we focus on the part of the free energy which depends on the orientational order parameters, that is,
\begin{equation}
\label{Eq:19}
\beta \mathcal{F}^{or} / V = \beta \mathcal{F}(\{\mathbf{Q}\}) / V - (\mathcal{F}_0^A + \mathcal{F}_0^B + \mathcal{F}_0^{AB})\,,
\end{equation}
where the terms on the right side are defined in equations~\eref{Eq:16} and~\eref{Eq:17a}--\eref{Eq:17b}, respectively. We further specialize on situations
where the hard spherocylinders have equal diameters, i.e., $d_A = d_B = d$, but different aspect ratios, $\kappa_A \neq \kappa_B$. This allows
to introduce a dimensionless form of the orientational free energy, $\mathscr{F}^{or}=\beta \mathcal{F}^{or} d^3/V$, and the reduced densities $\rho_i^* = \rho_i d^3$. 
 
With these assumptions, the orientational free energy of a binary mixture of hard spherocylinders becomes
\begin{eqnarray}
\label{Eq:20}
\fl \mathscr{F}^{or} &= \mathscr{F}^{or}(\{\rho_i^* , \kappa_i\};\{\mathbf{Q}^i\})\nonumber\\
\fl &= \sum_{i=A,B} \Big\lbrace A_i(\rho_i^* , \kappa_i) (\mathbf{Q}^i:\mathbf{Q}^i) - B_i(\rho_i^* ) \Tr(\mathbf{Q}^i \cdot \mathbf{Q}^i \cdot \mathbf{Q}^i) + C_i(\rho_i^* , \kappa_i) (\mathbf{Q}^i:\mathbf{Q}^i)^2 \Big\rbrace \nonumber \\
\fl & \qquad - {A_{AB}(\{\rho_i^*\},\{\kappa_i\})}(\mathbf{Q}^A:\mathbf{Q}^B)  - {C_{AB}(\{\rho_i^*\},\{\kappa_i\})} (\mathbf{Q}^A:\mathbf{Q}^B)^2\,,
\end{eqnarray}
where the coefficients $A_i$, $B_i$, $C_i$, $A_{AB}$ and $C_{AB}$ are given by
\begin{eqnarray}
\label{Eq:21}
\fl A_i(\rho_i^* , \kappa_i) &= \frac{1}{2} \rho_i^* - \frac{1}{10} \rho_i^{*2} \left( \frac{5\pi^2}{64} - \frac{5\pi}{12}\kappa_i + \frac{5\pi}{16}\kappa_i^2  \right)G_{ii}\,,\\
\label{Eq:22}
\fl {B_i(\rho_i^* )}  &= \frac{\sqrt{30}}{21} \rho_i^*\,\\
\label{Eq:23}
\fl C_i(\rho_i^* , \kappa_i) &=\frac{19}{28} \rho_i^* - \frac{1}{18} \rho_i^{*2} \left( \frac{9\pi^2}{512} + \frac{\pi}{40}\kappa_i + \frac{9\pi}{128}\kappa_i^2 \right)G_{ii}\,,\\
\label{Eq:24}
\fl {A_{AB}(\{\rho_i^*\},\{\kappa_i\})} &=  {\frac{1}{5}\rho_A^* \rho_B^* \left( \frac{5\pi^2}{64} - \frac{5\pi}{24} (\kappa_A + \kappa_B) + \frac{5\pi}{32}\kappa_A\kappa_B \right) G_{AB}}\,,\\
\label{Eq:25}
\fl {C_{AB}(\{\rho_i^*\},\{\kappa_i\})} &=  {\frac{1}{9}\rho_A^* \rho_B^* \left( \frac{9\pi^2}{512} + \frac{\pi}{80} (\kappa_A + \kappa_B) + \frac{9\pi}{256}\kappa_A\kappa_B \right) G_{AB}}\,.
\end{eqnarray}
An explicit expression for the quantities $G_{ij}$ in equations~\eref{Eq:20}--\eref{Eq:23} 
results from a generalization of the Carnahan--Starling theory for HS systems to mixtures~\cite{Carnahan1969, Camp1996, Wensink2001}, yielding 
\cite{Malijevsky2008} 
\begin{eqnarray}
\label{Eq:26}
G_{ij} = \frac{4 - 3(\rho_i^* \nu_i + \rho_j^* \nu_j)}{4(1 - \rho_i^* \nu_i - \rho_j^* \nu_j)^2}\,,
\end{eqnarray}
where $\nu_i = v_i / d^3$.
\subsection{Equilibrium transition}
As a background for our later discussion of mixtures under shear we first investigate the isotropic--nematic transition in equilibrium. Specifically, we are interested in the stability
of the two phases in dependence of the densities
and aspect ratios. In equilibrium one expects a {\em uniaxial} nematic phase described by a single director, $\mathbf{n}$, characterizing the preferred alignment of the rods~\cite{Frenkel1989, Dingemans2006}. The second rank tensor then reduces to $\mathbf{Q}^i = \sqrt{\frac{3}{2}} q_i \langle \stl{ \vec{n} \vec{n}} \rangle$, where $q_i$ is the eigenvalue related to $\vec{n}$. 
For the products appearing in~\eref{Eq:20} it follows that $\Tr(\mathbf{Q}^i \cdot \mathbf{Q}^j) = q_i q_j$ and $\Tr(\mathbf{Q}^i \cdot \mathbf{Q}^i \cdot \mathbf{Q}^i) = q_i^3/{\sqrt{6}}$. 
By definition of the second rank tensor [see~\eref{Eq:06} with $k=2$] the eigenvalues $q_i$ are related to the well known Maier--Saupe order parameter $S_2$ via 
$q_i = \sqrt{5} S_2$~\cite{Hess2015}.

As a further simplification, we focus on the so--called "semi--dilute" regime corresponding to low densities. Physically, this implies 
that the particles have few contacts~\cite{Solomon2010} and that long--range hydrodynamic interactions are negligible~\cite{Mewis2012}.
On a more formal level, the low--density limit implies that the contact values of the pair correlation functions tend to one, and the same holds for their density averages.
Thus, $G_{ij} \rightarrow 1$. Still, as we will demonstrate below for the case of a one--component system, the 
mesoscopic theory provides reliable results (in comparison to particle--based simulations~\cite{Bolhuis1997}) in the range
$\kappa=20$--$40$. We also note that although the choice $G_{ij}= 1$ is the same as that in original Onsager theory~\cite{Onsager1949}, the present free energy functional 
is different due to the more sophisticated treatment of the excluded volume (see Appendix~A).

\subsubsection{Single component system}
\begin{center}
 \begin{figure}
 {\includegraphics[width=0.85\linewidth,keepaspectratio]{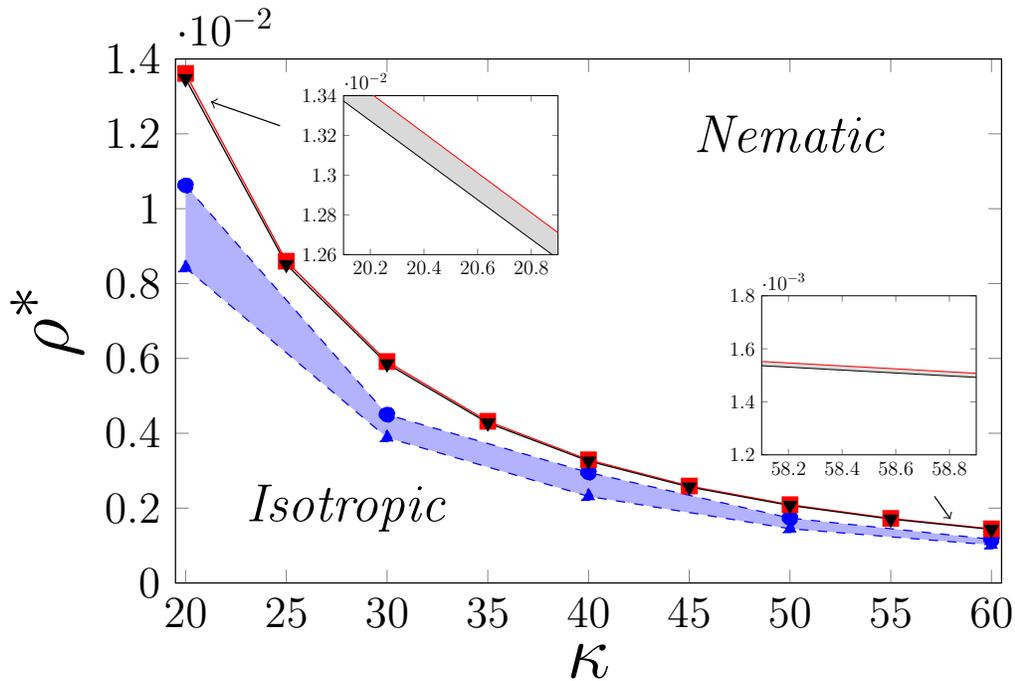}}
 \caption{(Color online) Isotropic--nematic transition of a system of hard spherocylinders with aspect ratios between $20$ and $60$. The (red) squares correspond to the lower limit of the density $\rho^*$ given by inequality~\eref{Eq:30}, while (black) down--triangles are the numerical solutions of the relation $A={3 B^2}/{64C}$. For comparison, the (blue) circles and up--triangles correspond to data for the isotropic--nematic phase coexistence from simulations of Bolhuis et al, see~\cite{Bolhuis1997}.}
 \label{Fig:01}
 \end{figure}
\end{center}

To validate our approach against literature results, we first consider the case of a one--component system. The orientational free energy then reduces to
the well--known Landau free energy expression~\cite{Gennes1993}
\begin{eqnarray}
\label{Eq:27}
\mathscr{F}^{or} = A(\rho^* , \kappa) q^2 - \frac{B(\rho^*)}{\sqrt{6}} q^3 + C(\rho^* , \kappa) q^4\,,
\end{eqnarray}
where the coefficients now depend on the aspect ratio and the density. From ~\eref{Eq:27}, we find the three stationary solutions
(determined by $d \mathscr{F}^{or}/d q=0$)
\begin{eqnarray}
\label{Eq:28}
q^0\,\, &= 0 \,,\\
\label{Eq:29}
q^{\pm}\, &= \frac{\sqrt{6} B}{16 C}\left(1 \pm \sqrt{1- \frac{64 AC}{3B^2}}\right)\,.
\end{eqnarray}
As usual, the isotropic phase becomes globally unstable (i.e., $d^2 \mathscr{F}^{or}/d q^2|_{q^0}<0$) when the second--order coefficient $A(\rho^* , \kappa)$ changes sign from positive and negative.
In terms of the density, this implies that within the stable or metastable isotropic phase,
\begin{eqnarray}
\label{Eq:30}
\rho^* < 5{\left( \frac{5\pi^2}{64} - \frac{5\pi}{12}\kappa + \frac{5\pi}{16}\kappa^2 \right)^{-1}}\,.
\end{eqnarray}
Directly at $A=0$, the nematic state already exists as a metastable state. The corresponding value of the order
parameter follows from ~\eref{Eq:29} as $q^{c} = {\sqrt{6}B}/{8C}$.
Further, from the second derivative evaluated at $q^+$ one finds that the nematic state becomes globally unstable ($d^2 \mathscr{F}^{or}/d q^2|_{q^{\pm}}<0$)
when $A>{3B^2}/{64C}$. The free energies $\mathscr{F}^{or}(q^0)$ and $\mathscr{F}^{or}(q^+)$ become equal at
\begin{equation}
\label{Eq:31}
A_{ni}=\frac{B^2}{24C}\,,
\end{equation}
The resulting order parameter at isotropic--nematic coexistence is given by
\begin{equation}
\label{Eq:32}
q_0 = \frac{\sqrt{6}B}{12C}\,.
\end{equation}
The stability ranges following from the above analysis are illustrated in the state diagram in~\fref{Fig:01}, where the axes are given by the density and the aspect ratio.
Specifically, the (red) $\blacksquare$  line in~\fref{Fig:01} represents the boundary of (meta--)stability of the isotropic state, 
determined by the maximal density fulfilling the inequality~\eref{Eq:30}, as function of $\kappa$. On the other hand, the (black) $\blacktriangledown$ line indicates the solutions of the relation $A={3 B^2}/{64 C}$ for a given value of $\kappa$. This 
line can be seen as the upper boundary of the stable isotropic state. 

For very large aspect ratios, $\kappa > 80$, our approach is consistent with Onsager's second virial theory~\cite{Onsager1949}.
For smaller aspect ratios, we compare our stability limits with the coexistence densities of isotropic and nematic phases (plotted in (blue) $\bullet$ and $\blacktriangle$) obtained by Bolhuis {\em et al} in 
\cite{Bolhuis1997} using Gibbs--ensemble simulations. We find that two sets of data agree reasonably well. Note that the orientational free energy ~\eref{Eq:27} can, by definition, 
not predict the coexistence {\em densities}, since we have neglected the densities $\rho_i$ as order parameters. 

\subsubsection{Binary mixture}

Depending on the overall density and concentrations, binary mixtures of particles with different aspect ratios can exhibit a very rich phase behavior~\cite{Purdy2005, Dennison2011}. 
Again, to calculate full phase diagrams, including first--order coexistence regions in density space, one would have to consider
the full free energy given in equations~\eref{Eq:02}--\eref{Eq:05}. 
Here we restricted ourselves to computing orientational state diagrams of the binary mixture by minimizing the orientational part of the free energy, $\mathscr{F}^{or}$, with respect to the uniaxial order parameters $q_A$ and $q_B$. The state diagrams are evaluated in the plane spanned 
by the total density $\rho^* = Nd^3/V$ and concentration $x_i = N_i/N$ (such that $\rho^*_i = N d^3 x_i / V = \rho^* x_i$ is the density of species $i$).
We focus on two mixtures characterized by aspect ratios also occurring in real systems~\cite{Purdy2005}.

\begin{center}
 \begin{figure*}
 \begin{tabular}{ c c }
 \subfloat[$\kappa_A = 20\,, \quad \kappa_B = 25\,.$]
 {\includegraphics[width=0.45\linewidth,keepaspectratio]{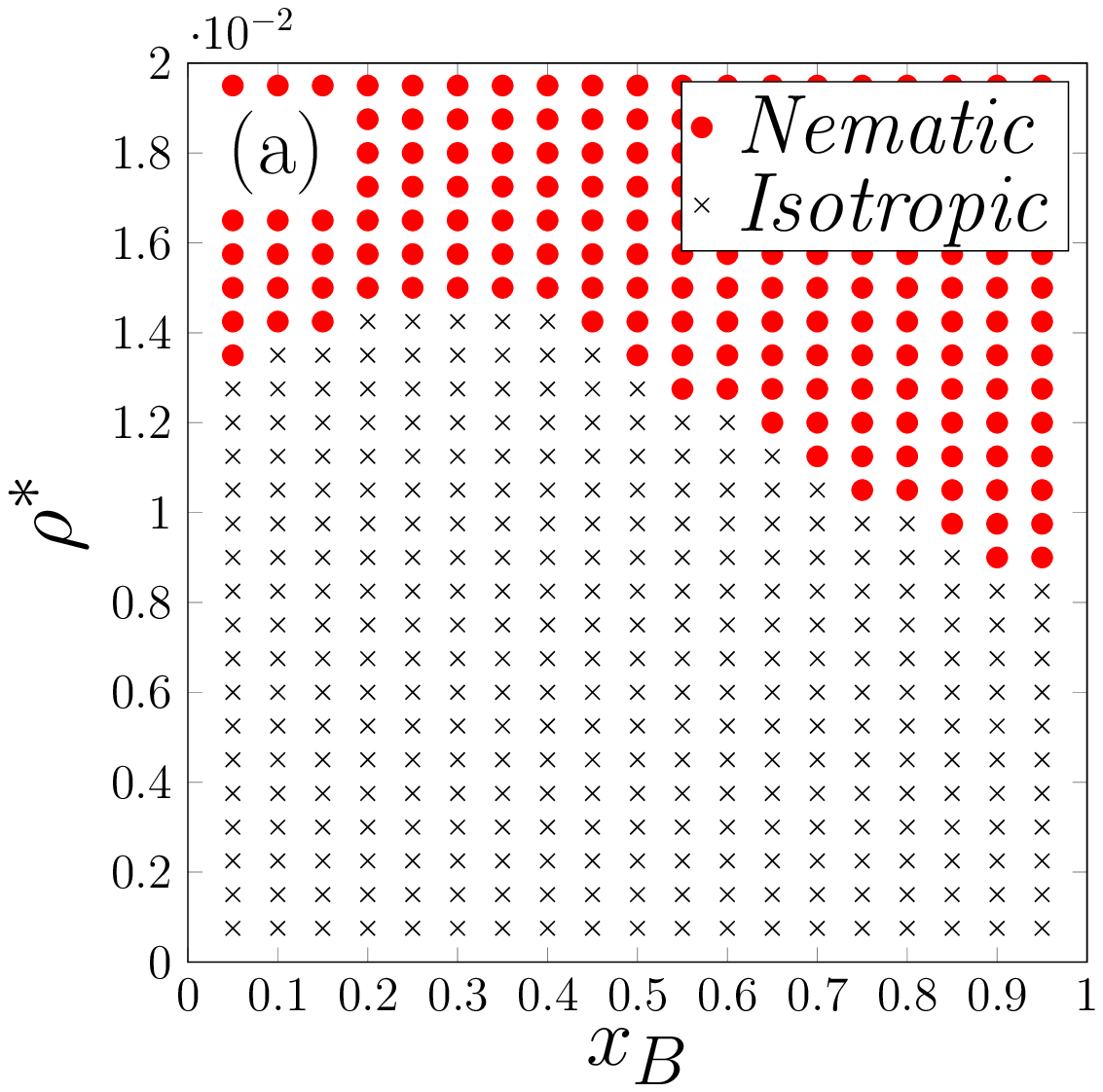}}
 & \subfloat[$\kappa_A = 24\,, \quad \kappa_B = 25\,.$]
 {\includegraphics[width=0.45\linewidth,keepaspectratio]{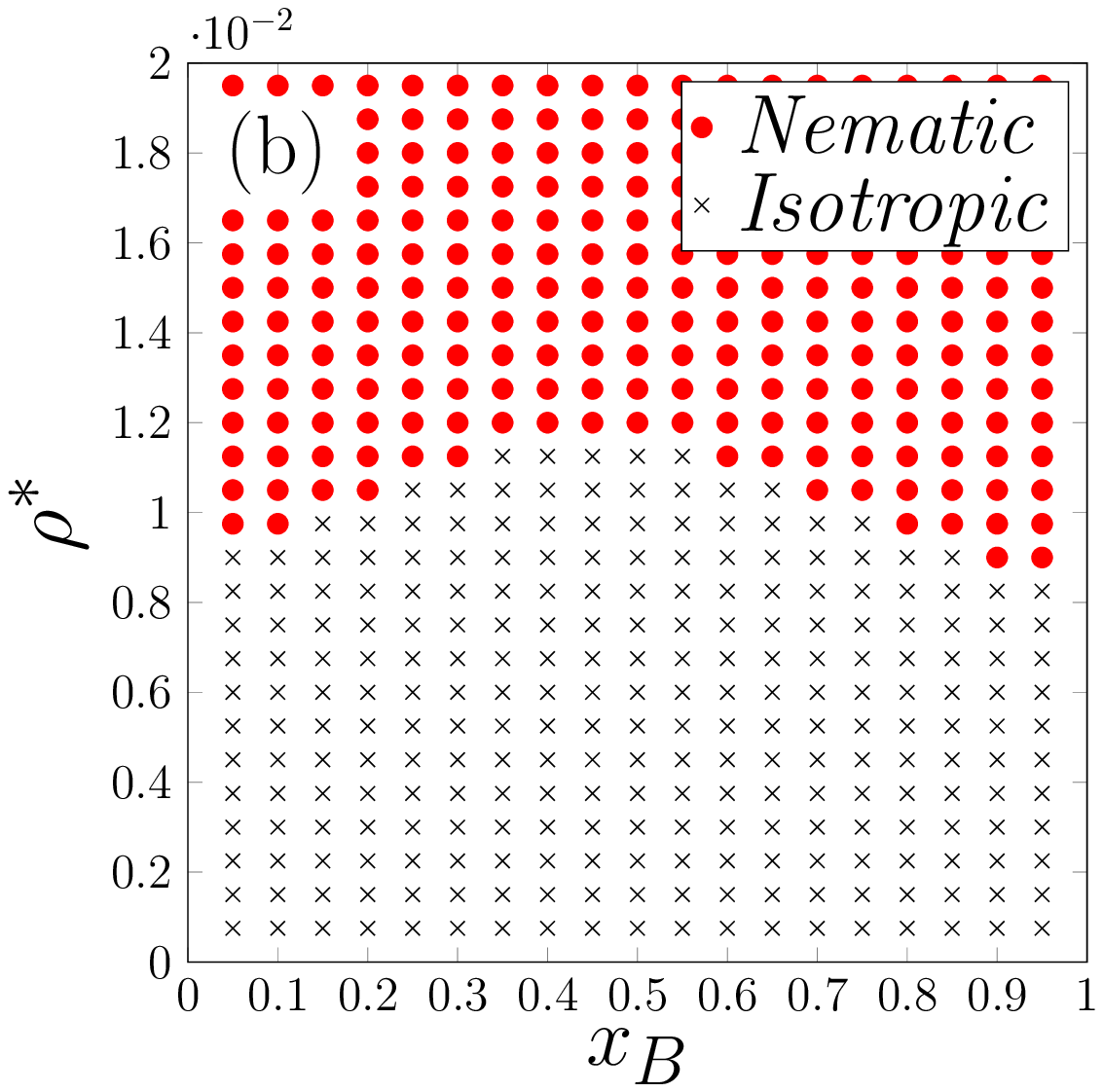}}
 \end{tabular}
 \caption{State diagrams for a mixture of long and short hard spherocylinders with uniaxial ordering. The nematic ordering is marked with (\textbullet), whereas  $(+)$ denotes an isotropic state. In the limit where $x_B = 0$ these diagrams coincide with the one component results portrayed in~\fref{Fig:01}.}
 \label{Fig:02}
 \end{figure*}
\end{center}

The methodology is the following. Using a standard Newton--Raphson minimization algorithm, we search for values of $q_A$ and $q_B$ corresponding to minima of $\mathscr{F}^{or}$. We then evaluate the orientational free energy at these points and determine the minimum value, $\mbox{min}\{\mathscr{F}^{or}\}$. In the case that $\mbox{min}\{\mathscr{F}^{or}\} = 0$ the isotropic phase is stable whereas if $\mbox{min}\{\mathscr{F}^{or}\} \neq 0$ a nematic state exists. The result of this procedure is shown in figures~\ref{Fig:02}(a--b).

In both diagrams, figures~\ref{Fig:02}(a) and~\ref{Fig:02}(b), we see that for relatively low densities, $\rho^* \sim 0.005$, the isotropic state is stable for all concentrations. As the total density of the system increases, the isotropic phase becomes unstable with respect to the nematic, and for sufficiently high densities, $\rho^* \sim 0.020$, the system is in the nematic state irrespective of the concentration.

As expected, the mole fraction of long spherocylinders plays an important role for the location of the transition. This is observed by focusing on a constant density value, $\rho^* \sim 0.0075$ in~\fref{Fig:02}(b). For $x_B = 0.2$ the nematic phase does not exist, but as $x_B$ increases, $x_B = 0.8$, the nematic state appears. In the limiting case where $x_B = 0$ the results coincide with the one component system presented in~\fref{Fig:01}.

Clearly, it would be very interesting to compare the equilibrium behavior found here to that of real systems. However, to our knowledge, experimental results
have only be reported
on binary mixtures of fd--viruses with {\em equal} length and {\em different} diameters~\cite{Purdy2005}. This is different from the mixture of rods
with different lengths and equal diameters considered in~\fref{Fig:02}. In principle, one could apply the present theory also to the first case by
rescaling the number densities via the length (instead of the diameter) of the particles and recalculating the excluded volume coefficients in equations~\eref{Eq:21}--\eref{Eq:25}.
Here we concentrate on the case of different length, the final motivation being the non--equilibrium behavior of rod--disk mixtures such as (models of) blood~\cite{Guu2013}.

\subsubsection{Free energy scaling}
For the subsequent analysis it is convenient to re--write~\eref{Eq:20} into a dimensionless form
using an appropriate scaling of the free energy. In principle, different scalings are possible. 
Here we follow the strategy used in earlier studies~\cite{Hess1975, Hess1976, Pardowitz1980, Hess2015} of one--component systems.

Specifically, we introduce the scaled alignment tensors, $\mathbf{\tilde{Q}}^i$ and the scaled free energy, $\mathscr{F}^{or}_{ref}$, as
\begin{equation}
\label{Eq:33}
\mathbf{\tilde{Q}}^i = \frac{\mathbf{Q}^i}{ q^i_0} \quad \mbox{and} \quad
\tilde{\mathscr{F}}^{or}= \frac{\mathscr{F}^{or}}{\mathscr{F}^{or}_{ref}}
\end{equation}
where $q^i_0$ is the coexistence value of the uniaxial order parameter of the corresponding one--component system given in~\eref{Eq:32},
and $\mathscr{F}^{or}_{ref}= 2\,C_{AB}\left(q^{A}_0 q^{B}_0\right)^2$. In the latter expression, $C_{AB}$ is the coupling coefficient of the fourth--order term $(\mathbf{Q}^A:\mathbf{Q}^B)^2$ of the free energy
~\eref{Eq:20}. 

To simplify the expression for the scaled free energy resulting from inserting ~\eref{Eq:33} into ~\eref{Eq:20}, we make the following assumptions.
First, we set the coefficients $B_i$ and $C_i$ [see equations~\eref{Eq:23} and~\eref{Eq:24}, respectively] equal to their (positive) values at I--N coexistence of the one--component system, that is,
\begin{equation}
\label{Eq:34}
B_i = B_i(\rho^*_{i_{ni}})  \quad \mbox{and} \quad C_i = C_i(\rho_{i_{ni}}^*,\kappa_i) \,, 
\end{equation}
where the densities $\rho^*_{i_{ni}}$ are calculated by solving ~\eref{Eq:31}.
Thus, we henceforth neglect changes of $B_i$ and $C_i$ with $\rho_i$. We further suppose that the fourth--order coefficient of the one--component system, 
$C_A$ and $C_B$, are related to $C_{AB}$ via 
\begin{eqnarray}
\label{Eq:35}
C_{AB} = C_A \left( \frac{q^A_0}{q^B_0}\right)^2 = C_B \left( \frac{q^B_0}{q^A_0}\right)^2 \,.
\end{eqnarray}
Equation~\eref{Eq:35} implies that for given values of $\kappa_i$ and $\rho_i^*$,  $|C_{AB}^2 - C_A C_B |= 0$. To test this assumption for a representative example,
we consider a binary mixture in the nematic state ($\rho^* = 0.002$) where $30\%$ of the particles have an aspect ratio $\kappa_A = 20$ and the others have an aspect ratio $\kappa_B = 25$.
In this case we find $|C_{AB}^2 - C_A C_B|  \approx 3.79 \times 10^{-7}$, which makes our ansatz plausible.

With the simplifications~\eref{Eq:34}--\eref{Eq:35}, the scaled version of the orientational free energy [see~\eref{Eq:20}] becomes
\begin{eqnarray}
\label{Eq:36}
 \fl \tilde{\mathscr{F}}^{or}[\mathbf{\tilde{Q}}^A, \mathbf{\tilde{Q}}^B] &=  \frac{\mathscr{F}^{or}[\mathbf{Q}^A, \mathbf{Q}^B]}{\mathscr{F}^{or}_{ref}} \nonumber \\
 &= \sum_{i=A}^B \Big\lbrace \frac{\Theta_i}{2}\!(\mathbf{\tilde{Q}}^i:\mathbf{\tilde{Q}}^i)\! - \sqrt{6}\, \Tr(\mathbf{\tilde{Q}}^i \cdot \mathbf{\tilde{Q}}^i \cdot \mathbf{\tilde{Q}}^i)\! + \frac{1}{2}(\mathbf{\tilde{Q}}^i:\mathbf{\tilde{Q}}^i)^2 \Big\rbrace \nonumber \\
 &\qquad -\frac{\Theta_{AB}}{2}(\mathbf{\tilde{Q}}^A:\mathbf{\tilde{Q}}^B) - \frac{1}{2}(\mathbf{\tilde{Q}}^A:\mathbf{\tilde{Q}}^B)^2\,.
\end{eqnarray}
In ~\eref{Eq:36}, the coefficients $\Theta_i$ of the second--orders term are given by
\begin{eqnarray}
\label{Eq:37}
\Theta_i = 24\frac{A_i(\rho^*_i,\kappa_i) C_i(\rho_{i_{ni}}^*,\kappa_i)}{B_i(\rho_{i_{ni}}^*)^2 }  = \frac{A_i(\rho^*_i,\kappa_i)}{A_{ni}(\rho_{i_{ni}}^*,\kappa_i)} \,,
\end{eqnarray}
which shows the explicit dependence of $\Theta_i$ on the composition (that is, on $\rho_i$) and $\kappa$. The remaining coefficient 
$\Theta_{AB}$ is a positive quantity which is depends not on the composition, but only on the aspect ratios. Explicitly,
\begin{eqnarray}
\label{Eq:38}
\Theta_{AB} = \frac{9\left[ \frac{5\pi^2}{64} - \frac{5\pi}{24} (\kappa_A + \kappa_B) + \frac{5\pi}{32}\kappa_A\kappa_B \right]}{5 \left[ \frac{9\pi^2}{512} + \frac{\pi}{80} (\kappa_A + \kappa_B) + \frac{9\pi}{256}\kappa_A\kappa_B \right] q_0^A q_0^B}\,.
\end{eqnarray}
In the limiting case of the one--component system, the scaled free energy~\eref{Eq:36} reduces to the corresponding expression given in~\cite{Hess1975, Hess1976, Pardowitz1980, Hess2015}.

\subsubsection{Analysis of the scaled free energy}
\begin{center}
 \begin{figure}
 {\includegraphics[width=0.85\linewidth,keepaspectratio]{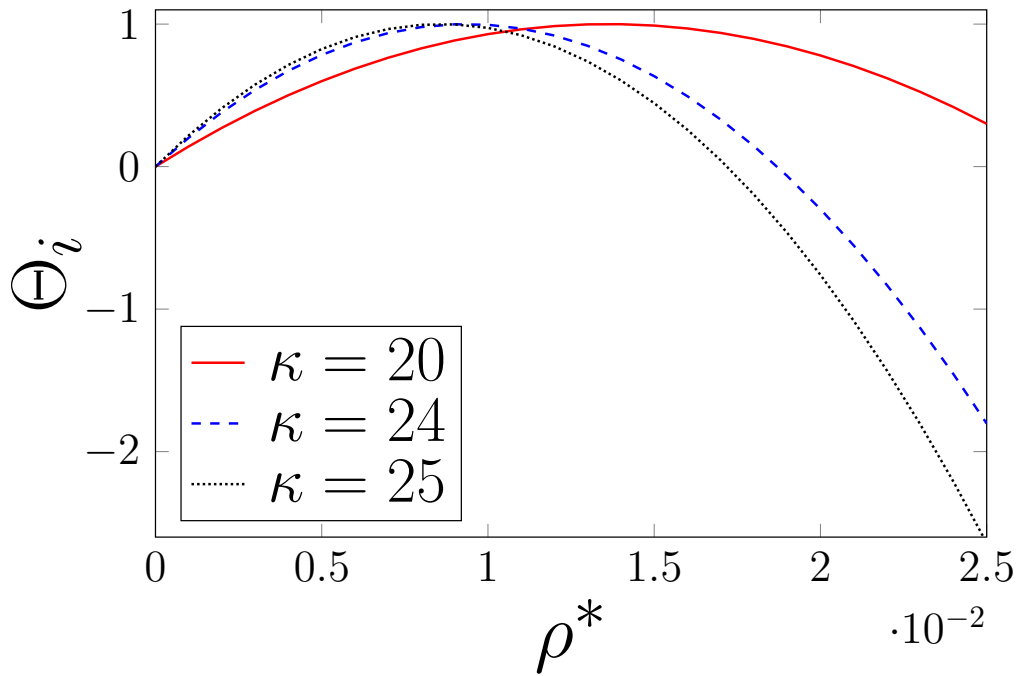}}
 \caption{Relation of $\Theta_i$ versus the reduced density, given by~\eref{Eq:37}, for equimolar binary mixtures $(x_A = x_B = 0.5)$ and different aspect ratio. The coexistence densities $\rho^*_{i_{ni}}$ are given by the roots of~\eref{Eq:31}.}
 \label{Fig:03}
 \end{figure}
\end{center}

\begin{center}
 \begin{figure}
 \begin{tabular}{ c c }
  {\includegraphics[width=0.45\linewidth,keepaspectratio]{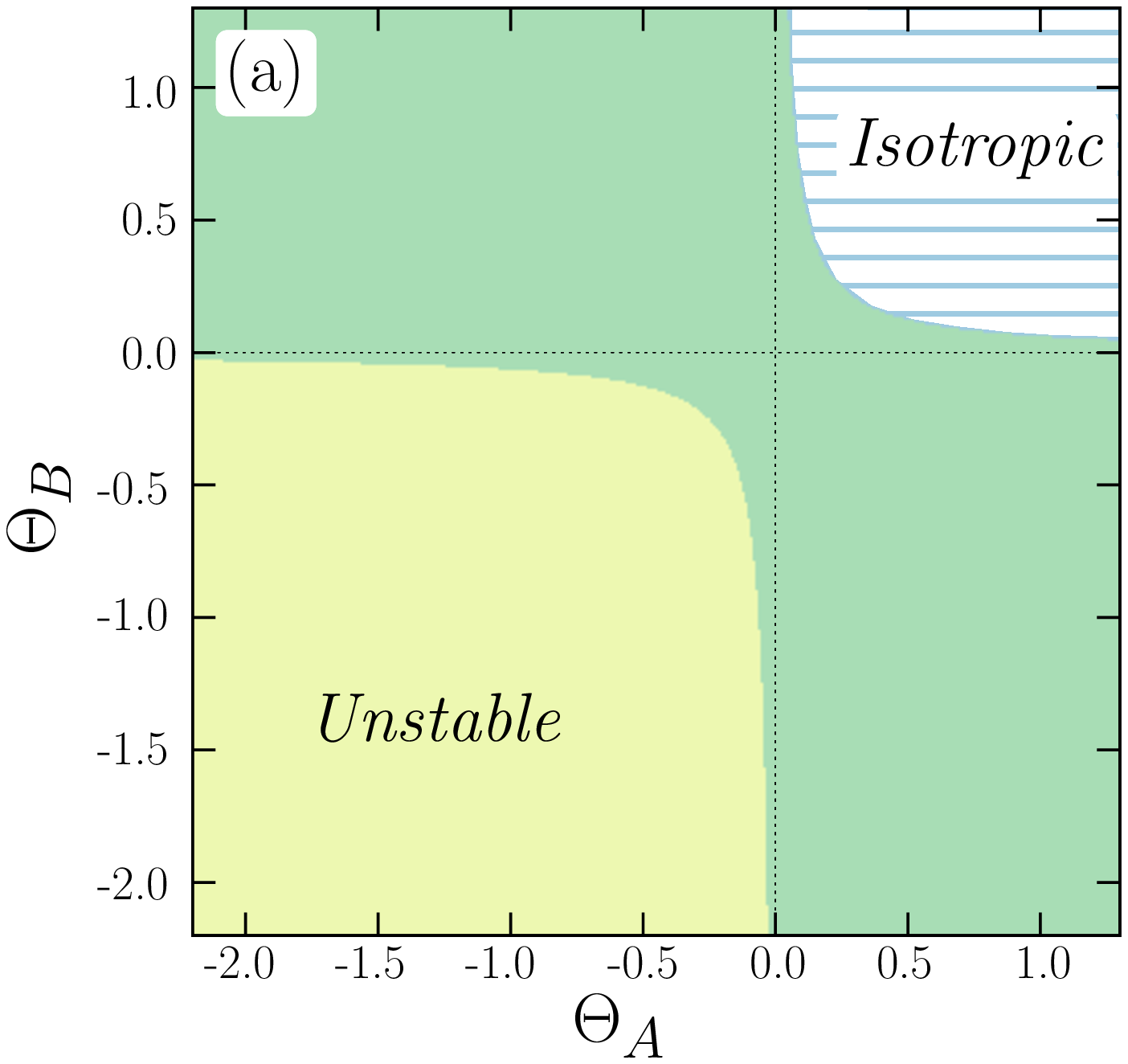}}
 &  {\includegraphics[width=0.45\linewidth,keepaspectratio]{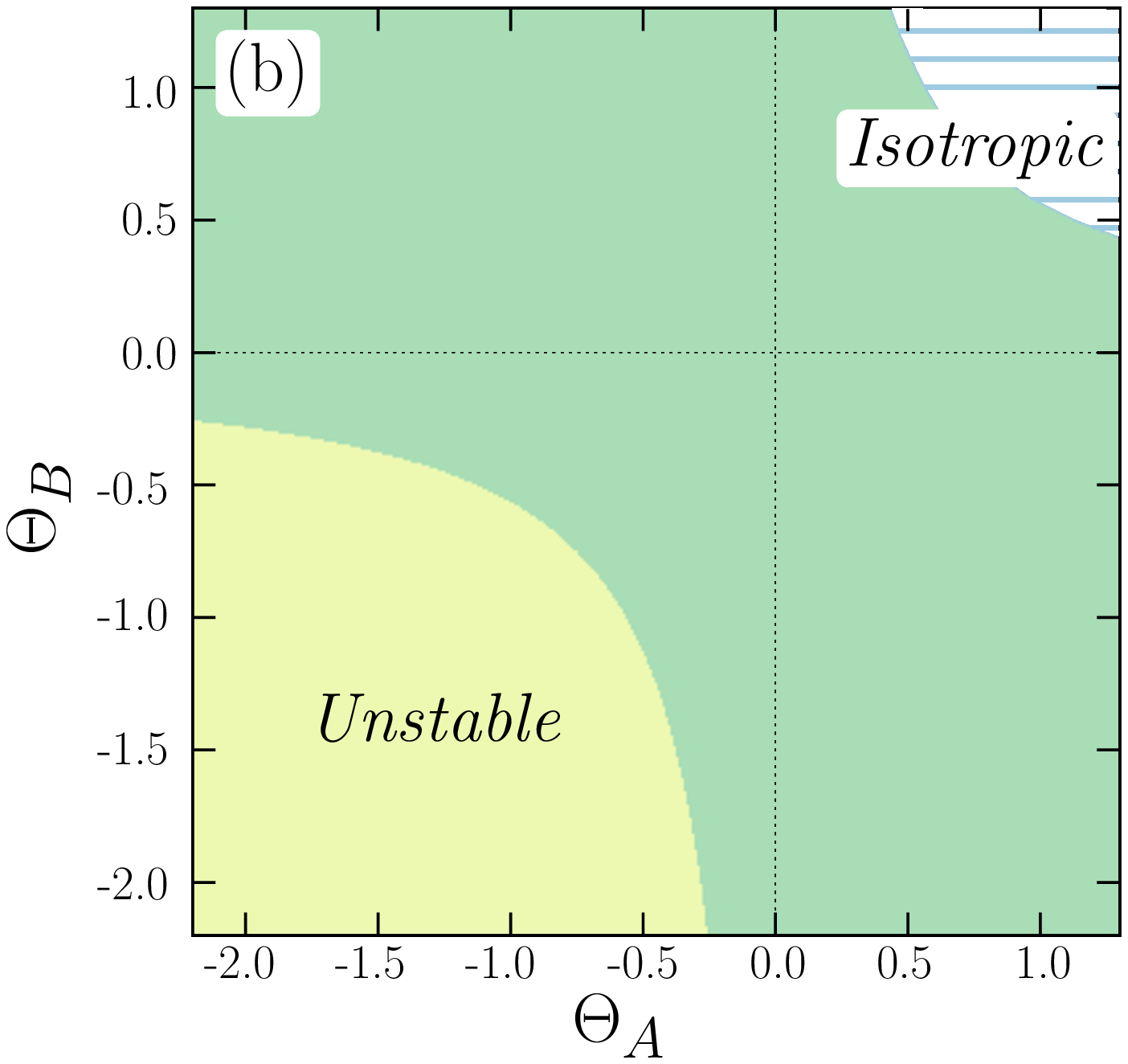}}
 \end{tabular}
  \caption{Stability regions of the isotropic state ($\mathbf{\tilde{q}}_{t} = (0,0)$) according to~\eref{Eq:41}. The values of $\Theta_{AB}$ are (a) $\Theta_{AB} = 0.50$ and (b)  $\Theta_{AB} = 1.50$.}
 \label{Fig:04}
 \end{figure}
\end{center}

To better understand the properties of the scaled free energy we consider, first, the explicit density dependence of the coefficients $\Theta_i$ appearing in front of the
quadratic powers of $\mathbf{\tilde{Q}}^i$ [see ~\eref{Eq:37}].
As an illustration, we show in~\fref{Fig:03} numerical results for an equimolar binary mixture and three values of the aspect ratio (note that each $\Theta_i$ depends only on $\kappa_i$).
One observes the same qualitative behavior irrespective of the actual value of $\kappa$: the coefficient $\Theta_i$ first increases with density and then reaches a maximum, after which it monotonically decreases and changes sign from positive to negative. This change of sign is expected (within the Landau picture) for a system displaying a phase transition.
The corresponding density is the smaller, the larger $\kappa$. 
To examine the relation between the sign of $\Theta_i$ and the mixture's stability, we consider the case of uniaxial order. In this case~\eref{Eq:36} reduces to
\begin{eqnarray}
\label{Eq:39}
\tilde{ \mathscr{F}}^{or}_{uni}= \sum_{i=A}^B  & \lbrace \frac{\Theta_i}{2} \tilde{q}_i^2 - \tilde{q}_i^3 + \frac{1}{2}\tilde{q}_i^4 \rbrace
 - \frac{\Theta_{AB}}{2} \tilde{q}_A \tilde{q}_B - \frac{1}{2} \tilde{q}_A^2 \tilde{q}_B^2 \,,
\end{eqnarray}
where $\tilde{q}_i = q_i/q^i_0$. The stability of a given minimum $\tilde{\bf q}$ can be determined from the Hessian matrix~\cite{Courant1989}
\begin{eqnarray}
\label{Eq:40}
   H(\tilde{q}_A,\tilde{q}_B)=\left(
                \begin{array}{cc}
                  \Theta_A - 6 \tilde{q}_A + 6 \tilde{q}_A^2 - \tilde{q}_B^2 & - \frac{\Theta_{AB}}{2} - 2 \tilde{q}_A\tilde{q}_B \\
                  &\\
                  - \frac{\Theta_{AB}}{2} - 2 \tilde{q}_A\tilde{q}_B & \Theta_B - 6 \tilde{q}_B + 6 \tilde{q}_B^2 - \tilde{q}_A^2 \\
                \end{array}
              \right)
\end{eqnarray} 
evaluated of the minimum. We now focus on the stability of the isotropic state,
$\mathbf{\tilde{q}}_t = (0,0)$. Stability then requires that the two eigenvalues or, alternatively, the two diagonal elements $\Theta_i$ and the determinant 
\begin{eqnarray}
\label{Eq:41}
\det[H(\mathbf{\tilde{q}}_{t})] = \Theta_A \Theta_B - \frac{\Theta_{AB}^2}{4}>0,
\end{eqnarray} 
are positive (where we recall that $\Theta_{AB}$ is a positive quantity).  An illustration of the regions of stability of the isotropic phase in the $\Theta_A$--$\Theta_B$ plane
at fixed $\Theta_{AB}$ is given in~\fref{Fig:04}. The larger $\Theta_{AB}$, the smaller is the region where the isotropic state is a minimum of the free energy.

It seems straightforward to repeat the same analysis for free energy minima corresponding to nematic states ($q_A\neq 0$, $q_B\neq 0$). However, in that context we face a problem 
already discussed in the context of one--component systems~\cite{Heidenreich2006, Heidenreich2009}, which concerns the {\em bounds} of possible values. This can be seen as follows:
The order parameters $q_A$ and $q_B$ are related to the Maier--Saupe order parameter $S_2$ by $q_i = \sqrt{5} S_2$~\cite{Hess2015}. The latter 
lies within the range $-1/2 \leq S_2 \leq 1$. The bounds of the corresponding scaled order parameters, ${q}_A$ and ${q}_B$, involve the factor $1/q^i_0$. Using~\eref{Eq:32} and the values of the coexisting densities $\rho^*_{i_{ni}}$ given by the roots of~\eref{Eq:31}, we find that 
the scaled order parameters are defined in the range
\begin{equation}
\label{Eq:42}
-1.30 \lesssim {q}_i \lesssim 2.60\,.
\end{equation}
However, free minimization of the scaled free energy~\eref{Eq:36} does not automatically respect these bounds; rather it yields, for a range of parameters, nematic minima outside the
allowed domain. To solve this problem, one could consider an "amended" version of the free energy, similar to what has been done for one--component systems
\cite{Heidenreich2006, Heidenreich2009}. However, this would go beyond the scope of the present work.

Another question, given the simplifications involved in our scaling of the free energy, is to which extent the stability conditions for the isotropic phase found here are consistent with
our earlier results obtained by minimization of the unscaled free energy (see~\fref{Fig:02}). As a check we have evaluated the quantities $\Theta_i$  for several state points in the diagrams plotted in~\fref{Fig:02}
The results are summarized in~\tref{Tab:01}, showing the consistency of unscaled and scaled free energy.
\begin{table}
\caption{\label{Tab:01}Numerical values of the constants $\Theta_A$, $\Theta_B$ and $\Theta_{AB}$ for equimolar binary mixtures $(x_A = x_B = 0.5)$. This values correspond to different points in the phase diagrams in figures~\ref{Fig:03}(a--b) and are calculated using~\eref{Eq:37} and~\eref{Eq:38}.}
\lineup
\begin{indented}
\item[] \begin{tabular}{  c c  c c  c c c}
\br
  & \multicolumn{2}{c}{\Fref{Fig:02}(a)} & \multicolumn{2}{c}{\Fref{Fig:02}(b)} & ~\Eref{Eq:41} & \\ [0.5mm]
  $\rho^*$  & \multicolumn{2}{c}{$\Theta_i $} &  \multicolumn{2}{c}{$\Theta_i $} & $det[H(\tilde{\mathbf{q}}_t)]$ & \\ [0.5mm]  
  $(10^{-2})$ &   $\kappa_A = 20$  & $\kappa_B = 25$ & $\kappa_A = 24$  & $\kappa_B = 25$ &  &\\[0.25mm] 
  \hline
  & & & & & & \\
  $0.50$ &  0.599 & 0.825  & 0.783  & 0.825  & + & {\em Isotropic} \\ [0.5mm]
  $1.00$ &  0.929 & 0.973  & 0.995  & 0.973  & + & {\em Isotropic} \\ [0.5mm]
  $1.75$ &  0.918 & \-0.073 & 0.239  & \-0.073 & - & {\em Unstable} \\ [0.5mm]
  $2.00$ &  0.779 & \-0.759 & \-0.298 & \-0.759 & - & {\em Unstable} \\ [0.5mm]
  & & & & & & \\
  \hline
  & & & & &\\
  & \multicolumn{2}{c}{$\Theta_{AB} = 1.203$ } & \multicolumn{2}{c}{$\Theta_{AB} = 1.223$ } &\\  [0.5mm]
  \br   
\end{tabular}
\end{indented}
\end{table}

\section{Shearing within the nematic state}
Based on the mesoscopic free energy functional derived in the preceding section we now explore the dynamical behavior of the system under shear. To this end
we use a straightforward extension of the Doi--Hess equations~\cite{Hess1975, Hess1976, Doi1978}, which may be derived either from the Fokker-Planck equation for the orientational
distribution function or via irreversible thermodynamics (for a recent formulation of the derivation, see~\cite{Hess2015}). The extension for binary mixtures involves 
two coupled equations for the order parameter tensors $\mathbf{Q}^A$ and $\mathbf{Q}^B$ in presence of a flow field $\mathbf{v}$.
Specializing on spatially homogeneous states the equations read
\begin{eqnarray}
\label{Eq:43}
\frac{d \mathbf{Q}^A}{dt} = 2\!\stl{{\mathbf{\Omega}} \cdot {\mathbf{Q}^A}}\!\!\! + 2\sigma\! \stl{{\mathbf{\Gamma}} \cdot {\mathbf{Q}^A}}\!\! - \frac{\mathbf{\Phi_A'}}{\tau_A}\! - \sqrt{2} \frac{\tau_{Ap}}{\tau_{A}} {\mathbf{\Gamma}}\,,\\
\label{Eq:44}
\frac{d \mathbf{Q}^B}{dt} = 2\!\stl{{\mathbf{\Omega}} \cdot {\mathbf{Q}^B}}\!\!\! + 2\sigma\! \stl{{\mathbf{\Gamma}} \cdot {\mathbf{Q}^B}}\!\! - \frac{\mathbf{\Phi_B'}}{\tau_{B}}\! - \sqrt{2} \frac{\tau_{Bp}}{\tau_{B}} {\mathbf{\Gamma}}\,,
\end{eqnarray}
where $\mathbf{\Omega} = (1/2)(\nabla \mathbf{v}\,^T - \nabla \mathbf{v})$ and $\mathbf{\Gamma} = (1/2)(\nabla \mathbf{v}\,^T + \nabla \mathbf{v})$ are tensors describing the
flow vorticity and flow deformation, respectively, and $\mathbf{\Phi}_i' $ are derivatives of the (unscaled) orientational free energy, i.e.
\begin{equation}
\label{Eq:45}
\mathbf{\Phi}_i' = \frac{ \partial \mathscr{F}^{or} ({\mathbf{Q}^A},{\mathbf{Q}^B})}{\partial \mathbf{Q}^i}.
\end{equation}

The degree of the perturbation exerted by the shear flow is measured through the prefactor of the last term of equations~\eref{Eq:43}--\eref{Eq:44},
that is, the ${\tau_{ip}}/{\tau_{i}}$, which defines the so--called tumbling parameter via  
$\lambda_i = -\tau_{ip}/\tau_{i}$. Both $\tau_i$ and $\tau_{ip}$ are phenomenological relaxation times dependent on the solvent's viscosity and the effective gyration radius of the colloidal particles~\cite{Hess1976}. It can be shown that their ratio, and thus, $\lambda_i$, is related to the particle aspect ratio~\cite{Hess1976} via
\begin{eqnarray}
\label{Eq:46}
\lambda_i = -\frac{\tau_{ip}}{\tau_{i}} = \sqrt{\frac{3}{5}}\frac{\kappa_i^2-1}{\kappa_i^2+1}\,.
\end{eqnarray}
For spherical particles $\lambda = 0$, whereas for prolate and oblate particles, $\lambda > 0$ and $\lambda < 0$, respectively. 

\subsection{Scaled dynamical equations}
In the following we specialize to a planar Couette shear flow characterized by the velocity $\mathbf{v} = \dot{\gamma} y \hat{e}_x$, where $\dot{\gamma}$ is the shear rate.
Further, we re--introduce the scaled alignment tensors $\mathbf{\tilde{Q}}^i$ and the scaled free energy according to~\eref{Eq:33}.
Equations~\eref{Eq:43} and~\eref{Eq:44} then become
\begin{eqnarray}
\label{Eq:47}
\frac{d \mathbf{\tilde{Q}}^A}{d \tilde{t}}  &= 2 \tilde{\dot{\gamma}}\!\stl{{\mathbf{\tilde{\Omega}}} \cdot {\mathbf{\tilde{Q}}^A}}\!\! + 2\sigma\tilde{\dot{\gamma}} \!\stl{\tilde{{\mathbf{\Gamma}}} \cdot {\mathbf{\tilde{Q}}^A}}\! - \tilde{\Phi_A'} + \sqrt{2}\tilde{\lambda_A} \tilde{\dot{\gamma}} {\mathbf{\tilde{\Gamma}}}\,,\\
\label{Eq:48}
\frac{d\mathbf{\tilde{Q}}^B}{d \tilde{t}}  &= 2 \tilde{\dot{\gamma}}\!\!\stl{{\mathbf{\tilde{\Omega}}} \cdot {\mathbf{\tilde{Q}}^B}}\!\! + 2 \sigma \tilde{\dot{\gamma}} \!\stl{{\mathbf{\tilde{\Gamma}}} \cdot {\mathbf{\tilde{Q}}^B}}\! - \tilde{\Phi_B'} + 	\sqrt{2}\tilde{\lambda_B}\tilde{\dot{\gamma}}{\mathbf{\tilde{\Gamma}}}\,,
\end{eqnarray}
where $\tilde{t} = t / t_{ref}$ and $\tilde{\dot{\gamma}} = \dot{\gamma} t_{ref} $ are the re-scaled time and shear rate, and the reference time connected to the phenomenological relaxation times $\tau_i$, 
is  $t_{ref}=\tau_i(q_0^i)^2{\mathscr{F}^{or}_{ref}}^{-1}$. The scaled vorticity and deformation tensors are $\mathbf{\tilde{\Omega}} = (1/2)(\hat{e}^x \hat{e}^y - \hat{e}^y \hat{e}^x)$ and 
$\mathbf{\tilde{\Gamma}} = (1/2)(\hat{e}^x \hat{e}^y + \hat{e}^y \hat{e}^x)$, respectively.
Further, the scaled tumbling parameters are given by 
\begin{equation}
\label{Eq:49}
\tilde{\lambda}_i = \lambda_i/q^i_0 =  \sqrt{\frac{3}{5}} \frac{1}{q^i_0} \frac{\kappa_i^2-1}{\kappa_i^2+1}.
\end{equation}
To give an example of typical values of the scaled tumbling parameters, we consider systems involving the PEG--coated {\em fd}--virus, like the ones used in the experiments of thick and thin colloidal rods of Purdy {\em et al.}~\cite{Purdy2005}. For these systems, the Maier-Saupe order parameter at
coexistence is $S_0 = \langle P_2\rangle \approx 0.5--0.8$~\cite{Purdy2003} . Using $q^i_0 = \sqrt{5} S_0$~\cite{Hess2015} one obtains $\tilde{\lambda}_i \sim 0.43--0.69$.
Finally, the derivative of the scaled free energy [see~\eref{Eq:36}] is given by
\begin{eqnarray}
\label{Eq:50}
\fl \Phi_i' = \Theta_i \mathbf{\tilde{Q}}^i - 2\sqrt{6} \stl{\mathbf{\tilde{Q}}^i \cdot \mathbf{\tilde{Q}}^i}  + 2 (\mathbf{\tilde{Q}}^i:\mathbf{\tilde{Q}}^i) \cdot \mathbf{\tilde{Q}}^i  - \frac{1}{2} \Theta_{AB} \mathbf{\tilde{Q}}^j - (\mathbf{\tilde{Q}}^A:\mathbf{\tilde{Q}}^B) \cdot \mathbf{\tilde{Q}}^j\,.
\end{eqnarray}
\subsection{Explicit equations of motion}
For a practical analysis of equations~\eref{Eq:47} and~\eref{Eq:48}, the second--order tensors $\mathbf{\tilde{Q}}^i$ are expanded in terms of a standard tensorial basis with five independent elements, see~\cite{Heidenreich2009, Rienaecker2002, Hess2015}. Introducing this tensor basis results into a system of ten coupled non--linear equations for $q_{0}^i, q_{1}^i, \cdots, q_{4}^i$, $i = A,B$. These
equations are given by
\begin{eqnarray}
\label{Eq:51}
  \dot{q}_0^i &= - (\phi_0^i + \varphi_{0}^j ) - \frac{\sqrt{3}}{3} \dot{\gamma}\sigma q_2^i\,,\nonumber \\
  \dot{q}_1^i &= - (\phi_1^i + \varphi_{1}^j ) + \dot{\gamma} q_2^i\,, \nonumber \\
  \dot{q}_2^i &= - (\phi_2^i + \varphi_{2}^j ) - \dot{\gamma} q_1^i - \frac{\sqrt{3}}{3}\dot{\gamma}\sigma q_0^i + \lambda_i \dot{\gamma} \,,\\
  \dot{q}_3^i &= - (\phi_3^i + \varphi_{3}^j ) + \frac{1}{2}\dot{\gamma} \left( \sigma + 1\right) q_4^i \,,\nonumber \\
  \dot{q}_4^i &= - (\phi_4^i + \varphi_{4}^j ) + \frac{1}{2}\dot{\gamma} \left( \sigma - 1\right) q_3^i \,.\nonumber
\end{eqnarray}
In these equations, the quantities $\phi_k^i$, which describe the relaxation of the particles towards equilibrium, are given by
\begin{eqnarray}
\label{Eq:52}
  \phi_0^i &= q_0^i \big( \Theta_i - 3 q_0^i + 2 \mathfrak{q}_i^{2} \big) + 3 (q_1^{i\,2} + q_2^{i\,2}) -\frac{3}{2}(q_3^{i\,2} + q_4^{i\,2})  \,,\nonumber \\
  \phi_1^i &= q_1^i \big( \Theta_i + 6 q_0^i + 2 \mathfrak{q}_i^{2} \big) - \frac{3}{2}\sqrt{3}\, (q_3^{i\,2} - q_4^{i\,2})\,,\nonumber  \\
  \phi_2^i &= q_2^i \big( \Theta_i + 6 q_0^i + 2 \mathfrak{q}_i^{2} \big) - 3\sqrt{3}\, q_3^{i} q_4^{i} \,,\\
  \phi_3^i &= q_3^i \big( \Theta_i - 3 q_0^i + 2 \mathfrak{q}_i^{2} \big) - 3\sqrt{3}\, (q_1^{i} q_3^{i} + q_2^{i} q_4^{i}) \,,\nonumber  \\
  \phi_4^i &= q_4^i \big( \Theta_i - 3 q_0^i + 2 \mathfrak{q}_i^{2} \big) + 3\sqrt{3}\, (q_1^{i} q_4^{i} - q_2^{i} q_3^{i})  \,.\nonumber
\end{eqnarray}
while the quantities $\varphi_{k}^i$, which take into account the inter--component interaction within the mixture, are given by
\begin{eqnarray}
\label{Eq:53}
  \varphi_{\alpha}^j = -q_\alpha^j  (\Theta_{AB}  + \mathfrak{q}_{AB}^{2})\,,
\end{eqnarray}
where $\alpha = 0,1,\cdots,4$. In equations~\eref{Eq:52}--\eref{Eq:53}, $\mathfrak{q}_i^{2} = \sum q_k^{i\,2}$ and $\mathfrak{q}_{AB}^{2} = \sum q_{k}^i q_{k}^j$. 

For one--component systems, it can be shown by general arguments~\cite{Strehober2013} (and has been demonstrated numerically~\cite{Rienaecker1999, Rienaecker2000}) 
that the structure of equations~\eref{Eq:51}--\eref{Eq:53} leads to various types of oscillatory solutions provided that the shear rate $\dot\gamma\neq 0$. 
In particular, for the case $\sigma=0$ (and appropriate thermodynamic state points, usually within the nematic state)
the equations predict the presence of two oscillatory "subunits" indicating oscillations within the shear plane ($q_1\neq 0$, $q_2\neq 0$, $q_3=q_4=0$) and
out of the shear plane ($q_3\neq 0$, $q_4\neq 0$), respectively. These oscillations have also been observed in numerous numerical studies~\cite{Hess2004, Hess2005}.
Here we follow these previous studies of one--component systems and set $\sigma=0$. Given the fact that our set of equations is just a direct extension of those
for one-component systems, we expect a variety of oscillatory solutions also in the mixture case.

To perform the numerical integration of equations~\eref{Eq:51}--\eref{Eq:53} we employ a standard Runge-Kutta algorithm with adaptive step size control~\cite{Press1992}. The starting values of the 
tensorial components $q_0^i,\cdots,q_4^i$ (where $i=A,B$) are chosen randomly in the interval $[0,1]$, using a uniform distribution. We note that the choice of initial values and the overall number of time--steps required to obtain reliable results (disregarding transient initial behavior) strongly depends on the chosen parameters. Therefore, all numerical integrations have been repeated several times.
To characterize the resulting dynamical states, we use an algorithm that recognizes the periodicity and sign change of the calculated time--dependent order parameters $q_k^i(t)$
\cite{Hess1994, Rienaecker1999, Rienaecker2002, Hess2004, Klapp2010, Strehober2013}.

\subsection{Numerical results}
In the following we investigate shear--driven systems whose equilibrium state ($\dot\gamma=0$) is nematic. Specifically, we set $\Theta_A = \Theta_B = -0.25$. According to the stability analysis in section 2.4.4,
the isotropic phase is then unstable irrespective of $\Theta_{AB}$. 

The section is divided into two parts. First, we discuss the effect of varying the shear rate, $\dot{\gamma}$, and the tumbling parameter of species B, $\lambda_B$, while $\lambda_A$ is fixed. 
We note that in real experiments, variation of the tumbling parameter is somewhat difficult since each value of this parameter corresponds to a different aspect ratio (and thus, different particle type),
see~\eref{Eq:49}. However,
in the framework of our dynamical equations, $\lambda_{B(A)}$ is the crucial parameter measuring the impact of the shear--induced perturbation, thus it seems worth to explore its role.
Specifically, we set $\lambda_A = 1.2=const$ and consider a range of values $\lambda_B>\lambda_A$, implying that the B--particles have larger aspect ratios.
The second part of the section then addresses the role of the "cross--coupling" parameter $\Theta_{AB}$.

\subsubsection{Variation of the tumbling parameter and the shear rate}
\begin{center}
\begin{figure*}
{\includegraphics[width=0.95\linewidth,keepaspectratio]{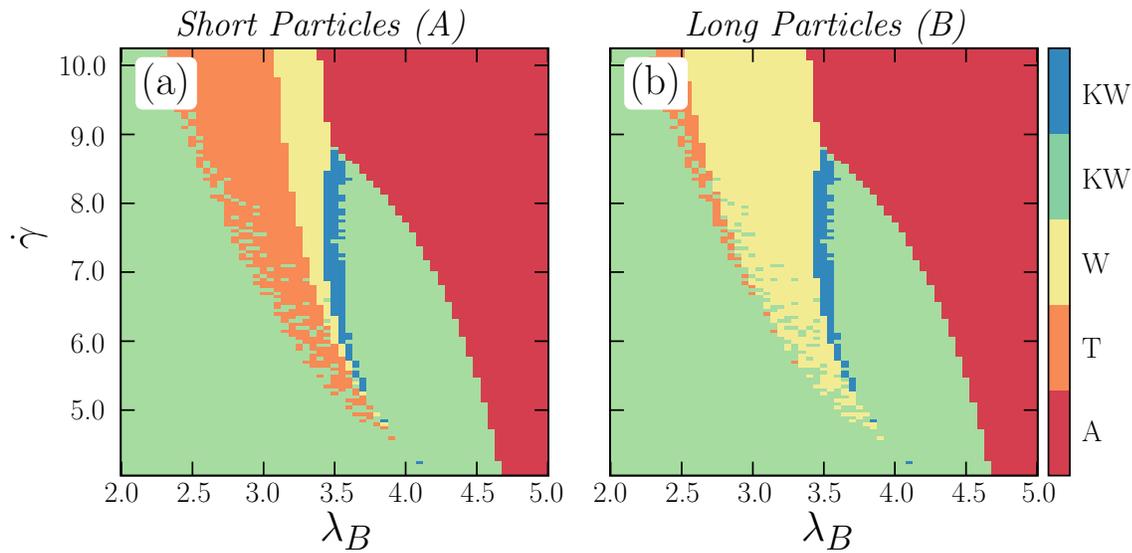}}
\caption{Dynamical state diagram for a binary mixtures characterized by $\Theta_A = -0.25$, $\Theta_B = -0.25$, $\Theta_{AB} = 1.50$, and $\lambda_A = 1.2$. Parts a) and b) illustrate the behavior
of the short and long particles, respectively. The areas are computed via direct numerical integration. Each colored region corresponds to different dynamical behavior (see color bar on the right side): A = Alignment, W = Wagging, T = Tumbling, KW = Kayak--Wagging, KT = Kayak--Tumbling.}
\label{Fig:05}
\end{figure*}
\end{center}

The numerical integration procedure described in section 3.2 leads to dynamical state diagrams for each of the two species which we present in~\fref{Fig:05}.
 Each colored region corresponds to a different dynamical state characterized by a specific behavior of 
 the respective nematic director, that is, the eigenvector
 related to the largest eigenvalue of the ordering tensor $\mathbf{Q}^{A(B)}$.
 Specifically, we observe
wagging states (W), tumbling states (T), kayaking--wagging states (KW), kayaking--tumbling states (KT), and (flow--)alignment states (A). The last
type of states corresponds to fixed points of the dynamical system, where
the director is "frozen" along a direction within the shear plane. All other states
mentioned so far are characterized by a time--dependent, regular oscillatory behavior 
of the coefficients $q_i(t)$ of $\mathbf{Q}(t)$. Physically, these oscillations correspond to oscillations
of the nematic director either {\it within} the shear plane (W, T), or {\it out of} the
shear plane (KW, KT). The boundaries between the colored regions in~\fref{Fig:05} correspond to different types of dynamical bifurcations. 
The simplest one is a (supercritical) Hopf bifurcation occurring at the boundary between alignment (A) and wagging (W). 
In~\cite{Strehober2013} we have presented a complete continuation analysis revealing the complexity of the
dynamical behavior already in the one--component case (where one has already five dynamical variables). 
Performing such an analysis for the mixture would have been beyond the scope for the present paper.
We therefore just note that the shape of each state diagram in~\fref{Fig:05} is qualitatively similar to that of one--component systems~\cite{Rienaecker1999, Rienaecker2000} sheared from a nematic equilibrium state, 
apart from the fact that the latter also
contain a chaotic region (C)~\cite{Rienaecker1999, Rienaecker2002, Rienaecker2000, Cates2002}. 
Here, we did not detect such a region
because our algorithm does not calculate Lyapunov exponents. Still, we expect that such a region also occurs in the mixture
system (for both species) at values of $\dot\gamma$, $\lambda_B$ where the regions A, KW and KT meet.

While the overall behavior of the two species is clearly very similar, there are interesting differences for specific values of $\dot\gamma$ and $\lambda_B$. 
As an example, consider the shear rate $\dot{\gamma}=7.5$ and $\lambda_B=3.0$. At these parameters, the long rods are in the W state while the short rods
are in a tumbling (T) state. In other words, both directors display oscillatory behavior {\em within} the shear plane, but with different characteristics:
Tumbling is characterized by full in--plane rotations of the nematic director, whereas wagging just implies finite back--and--forth--motion in angular space.

To quantify the differences, we have calculated the angle $\vartheta = \arccos[(\mathbf{d}^A \cdot \mathbf{d}^B)/(|\mathbf{d}^A||\mathbf{d}^B|)$
between the two nematic directors, $\mathbf{d}^A$ and $\mathbf{d}^B$.
For in--plane motion, the Cartesian components of (un--normalized) $\mathbf{d}^i$ are related (via a linear transformation~\cite{Klapp2010}) to the 
components $q^i_0, q^i_1, q^i_2$ of the full Q--tensors. Performing the scalar products between the directors then yields
\begin{eqnarray}
\label{Eq:54}
\vartheta(t) = \arccos \left( \frac{q^A_j q^B_j}{\sqrt{q^A_j q^A_j} \sqrt{q^B_j q^B_j}} \right)\,.
\end{eqnarray}
where $j=0,1,2$ and we used the Einstein convention. 

\begin{figure}
 \begin{tabular}{ c c }
 {\includegraphics[width=0.45\linewidth,height=0.4\linewidth]{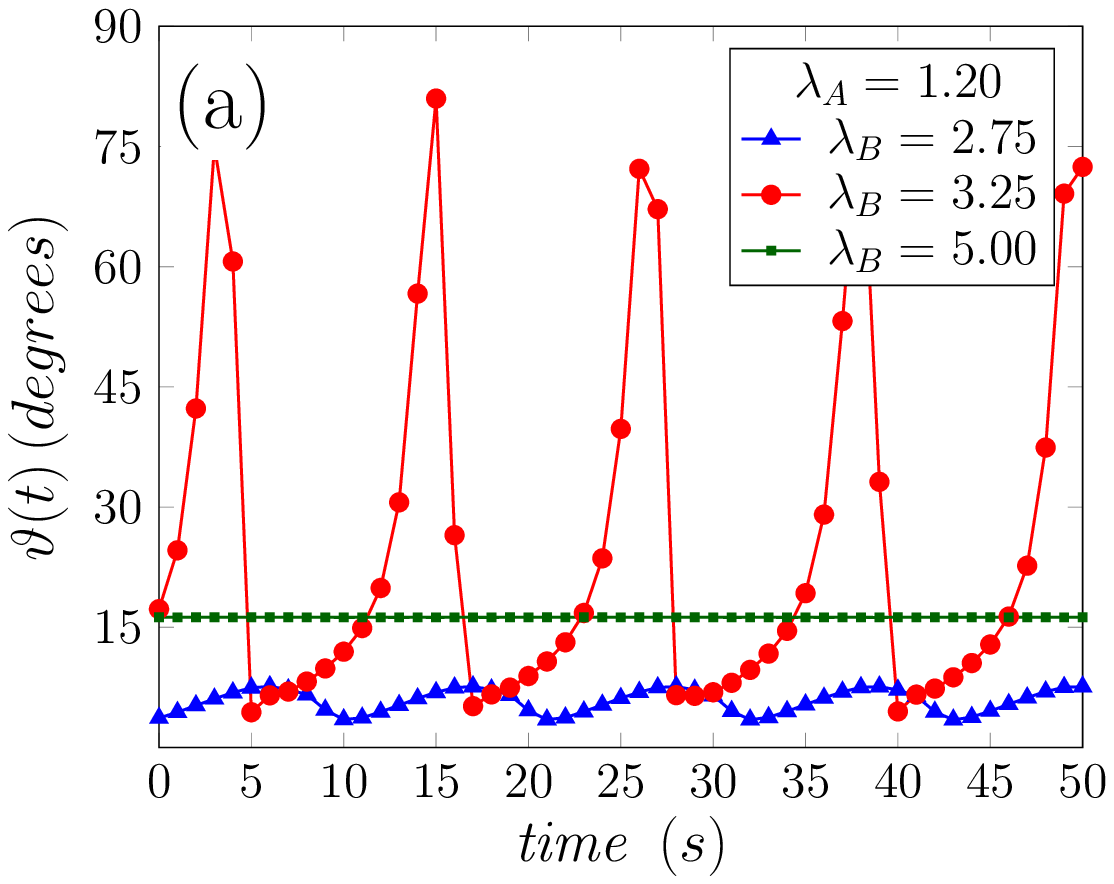}}
& {\includegraphics[width=0.45\linewidth,height=0.39\linewidth]{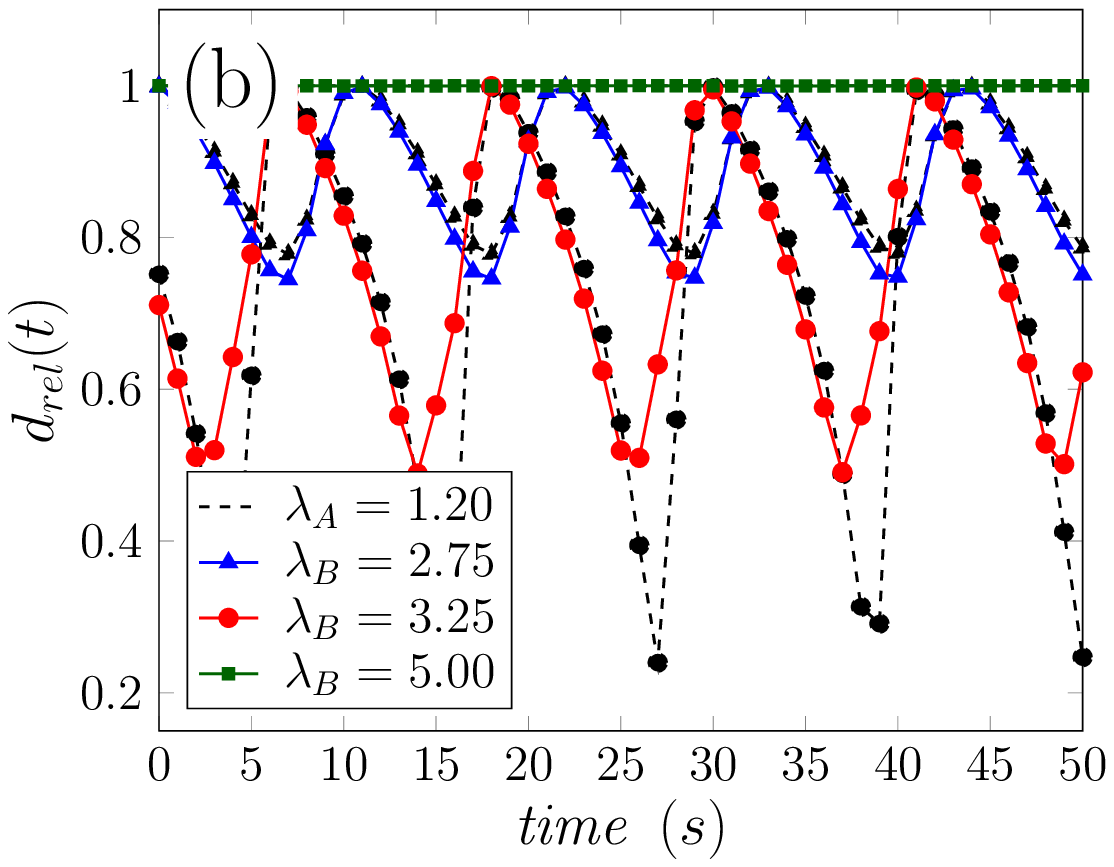}}
\end{tabular}
\caption{a) Angle between the nematic directors as function of time at
$\dot{\gamma} = 7.5$ and three values of $\lambda_B$. The green solid line ($\blacksquare$) corresponds to shear--aligned states of both species ($\lambda_B=5.0$),
whereas the blue line ($\blacktriangle$) corresponds to synchronized tumbling states ($\lambda_B=2.75$). Finally, the red line ($\bullet$) represents a case where the A--species (B--species) is
in a tumbling (wagging) state ($\lambda_B=3.75$). b) Relative alignment of the nematic directors as a function of time at $\dot{\gamma} = 7.5$. The dashed black lines and the solid lines correspond to the behavior of the A--species  ($\lambda_A=1.20$) and B--species (with varying $\lambda_B$), respectively. The ($\blacksquare$) symbol correspond to shear--aligned states of both components ($\lambda_B=5.0$) whereas the ($\blacktriangle$) correspond to synchronized tumbling states ($\lambda_B=2.75$). Finally, ($\bullet$) represents the coexisting T/W state ($\lambda_B=3.75$).}
\label{Fig:06}
\end{figure}

Numerical results at $\dot{\gamma}=7.5$ and three values of $\lambda_B$ are plotted in~\fref{Fig:06}(a), where we focus on times after the initial transient period.
At the largest $\lambda_B$ considered, both species are in a shear--aligned state. For single--component systems, it is well known~\cite{ Rienaecker2002, Rienaecker1999, Hess1994} that the shear--aligned state is characterized by a {\em finite} angle
between the (stationary) director and shear direction, and that this "flow angle" depends on the tumbling parameter. Therefore, one would expect the two directors of the mixture to have different 
(stationary) flow angles with respect to the shear direction, and consequently, to enclose a finite angle in between. This is exactly what we see from the corresponding curve in~\fref{Fig:06}(a). 
At $\lambda_B = 3.25$, both species are in the tumbling state (see~\fref{Fig:05}), with the directors displaying full rotations. Still, we see from~\fref{Fig:06}(a) 
that the angle {\em between} the tumbling directors is close to zero at all times considered. We interpret this behavior as (nearly) {\em synchronous} tumbling.
At $\lambda_B = 3.25$, however, the behavior is markedly different: Here we observe
a periodic variation of the angle between the directors, reaching values up to nearly $90$ degree. This reflects the simultaneous appearance of tumbling (species A) and wagging (species B).

To further compare the in--plane oscillations of the nematic directors in the different regions of the state diagram (see~\fref{Fig:04}) we also look at their relative amplitudes. To this end we introduce a quantity that compares, for each component of the mixture, the instantaneous magnitude $|\mathbf{d}^i|(t)$ with its maximum value $\mbox{max}\{|\mathbf{d}^i|\}$ (we recall at this point that the directors considered here
are {\em not} normalized to one). In terms of the components $q^i_0$, $q^i_1$, $q^i_2$, the relative alignment is therefore given by
\begin{eqnarray}
\label{Eq:55}
d^{A(B)}_{rel}(t) = \frac{\sqrt{q^{A(B)}_j q^{A(B)}_j}}{\mbox{max}\bigg\lbrace \sqrt{q^{A(B)}_j q^{A(B)}_j} \bigg\rbrace }  \, ,
\end{eqnarray}
where $j=0,1,2$ and we used the Einstein convention. This quantity is shown in~\fref{Fig:06}(b). As expected, in the shear--aligned state ($\lambda_B=5.0$)
the relative alignment of both components does not oscillate in time. Further, when both nematic directors are tumbling (i.e., perform full rotations, here at $\lambda_B = 2.75$), the oscillations are in phase and also their magnitude is comparable. In contrast, in the coexisting T/W state  ($\lambda_B = 3.25$), there is a marked difference between the amplitude of oscillations. Still, the oscillations are in phase. Therefore, we conclude that both, the T state and the coexisting T/W state are characterized by synchronous motion.

\subsubsection{Variation of the cross coupling}
\begin{center}
\begin{figure*}[ht!]
{\includegraphics[width=0.95\linewidth,keepaspectratio]{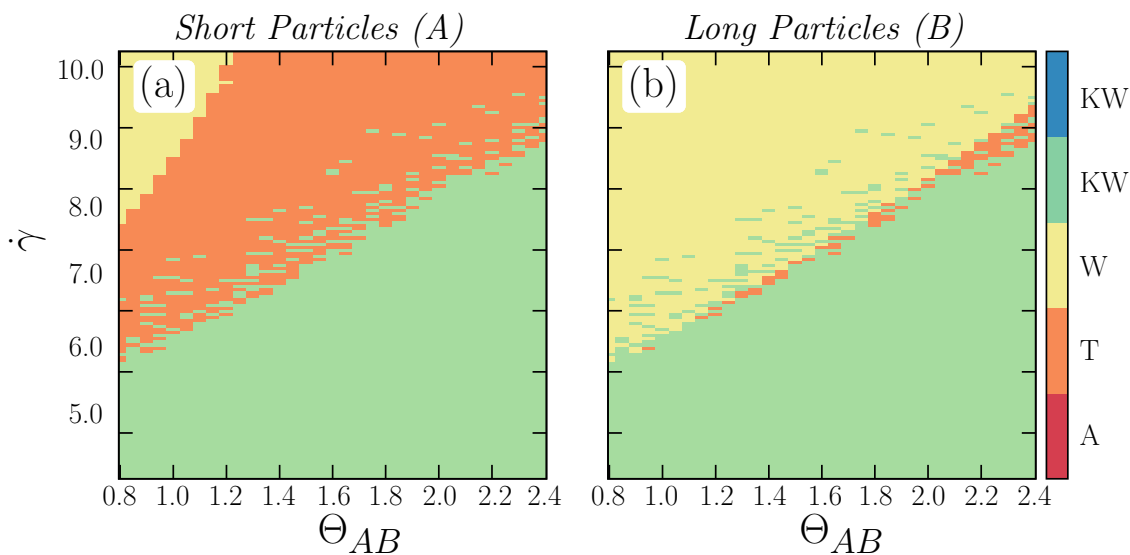}}
\caption{Dynamical state diagram in the plane spanned by shear rate and cross coupling parameter for a binary mixtures characterized by $\Theta_A = -0.25$, $\Theta_B = -0.25$, $\lambda_A = 1.2$ and $\lambda_B=3.0$. Parts a) and b) illustrate the behavior of the short and long particles, respectively.}
\label{Fig:07}
\end{figure*}
\end{center}

In view of the variety of dynamical states at {\em fixed} cross coupling parameter $\Theta_{AB}$ it is interesting to further explore the impact of this parameter.
To this end we present in~\fref{Fig:07} the state diagrams obtained at fixed tumbling parameters $\lambda_A = 1.20$ and  $\lambda_B = 3.0$, in the plane spanned
by $\Theta_{AB}$ and $\dot\gamma$. At $\dot\gamma=0$ (equilibrium), all systems considered are in the (meta--)stable nematic phase. 

At small shear rates $\dot\gamma\lesssim 5.0$, both species display an out--of--plane KT state for all cross coupling parameters considered. At larger shear rates the out--of--plane oscillations
then transform into in--plane oscillations of W or T type. As indicated from~\fref{Fig:07}, an increase of $\Theta_{AB}$ shifts this boundary between KT and T or W towards larger shear rates
for both components of the mixture.
A further effect becomes apparent when we consider a fixed, high shear rate, e.g., $\dot{\gamma}=8.0$. Here, an increase of $\Theta_{AB}$ can induce profound changes of the dynamical state, such 
as a change from W to T to KT. Along this way, there a large regions where the two components display {\em different} dynamical behavior (coexistence of T and W), 
similar to what we have discussed before.
Finally, we notice the small "islands" of KT regions within the T state (A--species) or W--state (B--species). This indicates a bi--stability of the solutions which may be interpreted
as a coexistence of in--shear--plane and out--of--shear plane oscillations (as already mentioned in~\cite{Strehober2013}). At this point, it would be very interesting to extend the present analysis
to spatially inhomogeneous systems: Indeed, as previously shown for a one--component system~\cite{Das2005, Chakraborty2010}, bistability in the homogeneous system can translate 
to spatially separated dynamical coexistence of different states, when one allows for inhomogeneity.

\section{Concluding remarks}
In this paper we have combined a microscopic free energy density  
functional~\cite{Malijevsky2008} and mesoscopic (continuum) equations
to investigate the orientational dynamics of binary mixtures of  
rod--like particles under planar (Couette) shear flow.
To establish the link between the microscopic and mesoscopic parts, we  
have performed an expansion
of the (Parson-Lee) functional in powers of the $\mathbf{Q}$--tensors  
of each species. This yields
a Landau--like expression for the orientational part of the free  
energy, the difference to the traditional LdG approach
being that the coefficients depend on the details of the  
pair interaction such as the aspect ratio.
The resulting functional enters via the relaxational term in the  
Doi--Hess equations for the $\mathbf{Q}$--tensors.

To explore the mixture's behavior in  thermodynamic equilibrium (zero  
shear), we have presented several calculations focusing on the  
stability of 
isotropic and nematic states. Indeed, experience with one--component  
systems~\cite{Strehober2013} shows that the nature of the  
underlying equilibrium state
is crucial for the dynamical behavior under shear. The equilibrium  
mixtures considered here were characterized by particles with equal  
diameters and different length.
This is somewhat in contrast to other recent studies focusing on  
"thick--thin" mixtures of particles with equal length and different  
diameters~\cite{Purdy2005}.
However, as noted at the end of section 2.4.2, the present approach  
can be easily modified towards thick--thin mixtures by a simple  
rescaling of densities.
A more severe restriction of our equilibrium investigations is that we  
have concentrated on the orientational part of the free energy alone.  
Thus,
our stability analysis does not give any information about phase  
transitions (or coupled phase transitions) involving the number  
densities as order parameters, an example
being macroscopic {\em demixing}. This would clearly be an important  
extension of the present work.

As a second step, we have explored the shear--induced non--equilibrium  
behavior starting from a globally nematic state. Similar to  
one--component systems,
we observe a variety of oscillatory dynamical states of the nematic  
directors, either with in--plane or with out--of--plane symmetry, as  
well as steady states.
One interesting new feature occurring for the binary mixture
is the appearance of states where the two directors perform {\em  
different} oscillatory motion. In particular, we observe a  
simultaneous appearance of
tumbling motion (short particles) and wagging motion (long particles).  
However, although the actual motion is different,
the time--dependence of quantities such as the angle between the  
directors reflects that the oscillations are {\em in synchrony}.

Another interesting result is the strong dependence of the dynamical  
state diagrams on the coupling parameter $\Theta_{AB}$ determined by  
the interactions between different species.
For example, by increasing $\Theta_{AB}$ at fixed shear rate, one can  
find transitions from in--plane oscillations to out--of--plane  
oscillations (kayaking--tumbling), either of same
or different type for the two components considered. In that sense,
$\Theta_{AB}$ has a similarly important role as the tumbling parameter  
$\lambda$. This prediction may be particularly interesting for  
experiments.

Given the appearance of mixed ("coexisting") dynamical states, it  
would be very interesting to extend the present analysis to spatially  
inhomogeneous systems, thereby allowing
for spatial domains characterized by different dynamical behavior.  
 From the methodological point of view, a straightforward strategy to  
include inhomogeneity into the equilibrium part
is to perform a gradient expansion
of the mesoscopic free energy functional. Within the dynamical part,  
one should then take into account the feedback effect caused by  
inhomogeneous orientational order parameters
on the velocity profile of the flow (see, e.g.,~\cite{Heidenreich2009}). A further route of research  
concerns
the rheological behavior of the mixture and its "tunability", e.g. via  
the aspect ratios. Work in these directions is in progress.
\section*{Acknowledgments}
We gratefully acknowledge financial support from the Deutsche Forschungsgemeinschaft (DFG) through the Research Training Group 1558, project B3.

\appendix

\section{Coefficients of the excluded volume of hard spherocylinders}
\label{ExcludedVolume}

The excluded volume of two spherocylinders, whose particle axes enclose the angle $\gamma$, is given in~\eref{Eq:18}.
In the following we aim at deriving analytic expressions for the coefficients 
$V_k^{ij}$, $k = 0,2,4$, appearing in the expansion of $V_{exc}(\gamma)$ in Legendre polynomials [see equations~\eref{Eq:09}--\eref{Eq:10}]. To this
end we expand the functions $ |\cos\gamma| $ and $E(\sin\gamma)$ in~\eref{Eq:18} 
up to fourth order in $\sin\gamma$, yielding
\begin{eqnarray}
\label{Eq:Ap01}
|\cos \gamma| &= (1 - \sin \gamma)^{1/2}\nonumber\\
&= 1 - \frac{1}{2}\sin^2 \gamma - \frac{1}{8}\sin^4 \gamma + \mathcal{O}(6)\,,
\end{eqnarray}
and
\begin{equation}
\label{Eq:Ap02}
E(\sin \gamma) = \frac{\pi}{2}  - \frac{\pi}{8}\sin^2 \gamma - \frac{3\pi}{128}\sin^4 \gamma + \mathcal{O}(6)\,.
\end{equation}
Neglecting the term $\mathcal{O}(6)$, the excluded volume becomes
\begin{eqnarray}
\label{Eq:Ap03}
\fl V_{exc}^{O(4)}(\gamma) = & \frac{\pi}{4} c_1 + \frac{\pi}{4} c_2 \sin\gamma 
  + \frac{\pi}{4} c_3 \left( 1 - \frac{1}{2}\sin^2 \gamma - \frac{1}{8}\sin^4 \gamma \right) \nonumber\\
 & + c_4 \left( \frac{\pi}{2}  - \frac{\pi}{8}\sin^2 \gamma - \frac{3\pi}{128}\sin^4 \gamma \right) + c_5 \sin\gamma\,,
\end{eqnarray}
where the constants $c_i$ depend on the aspect ratio and diameters of the particles and are given by
\begin{eqnarray}
\label{Eq:Ap04}
 c_1 = \kappa_i d_i^3 + \kappa_j d_j^3\,,&\quad c_2 = d_id_j(d_i + d_j)\,,\nonumber\\
 c_3 = \kappa_i d_i d_j^2 + \kappa_j d_j d_i^2\,, &\quad c_4 = (\kappa_i d_i + \kappa_j d_j) d_i d_j\,\nonumber\\
 c_5 = \kappa_i \kappa_jd_id_j (d_i + d_j)\,.
\end{eqnarray}
We note that the coefficients $c_3$ and $c_4$ vanish within the original Onsager approach~\cite{Onsager1949}.

Starting from~\eref{Eq:Ap03}, we can now employ~\eref{Eq:10} to calculate the expansion coefficients of interest ($k=0,2,4$):
\begin{equation}
\label{Eq:Ap05}
 V_k^{ij} = \frac{2k+1}{2} \int_0^\pi d\gamma\,\,V_{exc}^{O(4)}(\gamma) P_k(\cos\gamma)\sin\gamma\,.
\end{equation}

Using $P_0(\cos\gamma)=1$ and performing the angular integral, we obtain the zero--order coefficient 
\begin{equation}
\label{Eq:Ap06}
V_0^{ij} = \frac{\pi}{4}c_1 + \frac{\pi^2}{16}c_2 +\frac{3\pi}{20}c_3 +\frac{97\pi}{240}c_4 + \frac{\pi}{4}c_5\,.
\end{equation}
To calculate $V_2^{ij}$ we use that $P_2(\cos\gamma) = \frac{1}{2}(3 \cos^2 \gamma -1)$. Since
\begin{equation}
\label{Eq:Ap07}
 \int_0^\pi d\gamma\,\, P_2(\cos\gamma)\sin\gamma = 0\,,
\end{equation}
all linear functions of $\sin\gamma$ do not contribute to the coefficient. However, powers of $\sin\gamma$ do contribute since
\begin{eqnarray}
\label{Eq:Ap08}
 \int_0^\pi d\gamma\,\,\sin\gamma \, P_2(\cos\gamma)\sin\gamma &= -\frac{\pi}{16}\,, \\
\label{Eq:Ap09}
 \int_0^\pi d\gamma\,\,\sin^2\gamma \, P_2(\cos\gamma)\sin\gamma &= -\frac{4}{15}\,, \\
\label{Eq:Ap10}
 \int_0^\pi d\gamma\,\,\sin^4\gamma \, P_2(\cos\gamma)\sin\gamma &= -\frac{32}{105}\,.
\end{eqnarray}
Inserting these results into~\eref{Eq:Ap05} with $k=2$ we obtain
\begin{equation}
\label{Eq:Ap11}
 V_2^{ij} = -\frac{5\pi^2}{128}c_2 + \frac{3\pi}{28}c_3 + \frac{17 \pi}{168}c_4 - \frac{5\pi}{32}c_5.
\end{equation}
The coefficients $V_4^{ij}$ involve the Legendre polynomial $P_4(\cos\gamma) = \frac{1}{8}(35 \cos^4 \gamma - 30 \cos^2 \gamma + 3)$. 
We find that integrals involving linear and cubic terms in $\sin\gamma$ vanish
whereas the other integrals give non-zero results. Specifically,
\begin{eqnarray}
\label{Eq:Ap12}
 \int_0^\pi d\gamma\,\,\sin\gamma \, P_4(\cos\gamma)\sin\gamma &= -\frac{\pi}{128}\,, \\
\label{Eq:Ap13}
 \int_0^\pi d\gamma\,\,\sin^4\gamma \, P_4(\cos\gamma)\sin\gamma &= \frac{16}{315}\,.
\end{eqnarray}
With this we finally obtain
\begin{equation}
\label{Eq:Ap14}
 V_4^{ij} = -\frac{9\pi^2}{1024}c_2 - \frac{\pi}{140}c_3 - \frac{3 \pi}{560}c_4 - \frac{9\pi}{256}c_5\,.
\end{equation}
It is worth noting that, since the aspect ratio and the diameter are always positive, the coefficients $c_i$ must be positive as well. It follows 
that $V_4^{ij}$ is always negative, whereas the sign of $V_2^{ij}$ depends on the combination of sizes. This is different from standard
Onsager theory where $c_3=c_4=0$ and thus, $V_2^{ij}$ is always negative.

\section*{References}
\bibliographystyle{iopart-num}
\bibliography{References}

\end{document}